\documentclass[graybox]{svmult}

\usepackage{mathptmx}       
\usepackage{helvet}         
\usepackage{courier}        
\usepackage{type1cm}        
\usepackage{makeidx}         
\usepackage{graphicx}        
\usepackage{multicol}        
\usepackage[bottom]{footmisc}
\usepackage{hyperref}        
\usepackage{soul}            
\hypersetup{colorlinks=true,urlcolor=blue}
\usepackage[square,numbers,sort&compress,authoryear]{natbib}

\usepackage[squaren]{SIunits}
 \usepackage{mathtools}

\makeindex             

\definecolor{dark-red}{RGB}{238,0,3}
\definecolor{dark-blue}{RGB}{9,71,171}
\definecolor{dark-green}{RGB}{17,149,17}
\definecolor{deep-purple}{RGB}{165,0,220}
\definecolor{dark-yellow}{RGB}{175,175,0}

\newcommand{\dd}{\mbox{d}}
\newcommand{\MIP}{\mbox{MIP}}
\newcommand{\track}{\mbox{\footnotesize track}}
\newcommand{\annee}{\mbox{year}}
\newcommand{\gfrac}[2]{\displaystyle\frac{#1}{#2}}
\newcommand{\eff}{\mbox{\footnotesize eff}}
\newcommand{\scint}{\mbox{\footnotesize scint}}

\begin{document}


\title*{Time projection chambers for gamma-ray astronomy}

\author{Denis Bernard\thanks{corresponding author}, Stanley D. Hunter, Toru Tanimori }

\institute{Denis Bernard \at LLR,
Ecole polytechnique, CNRS / IN2P3 and IPP, 91128 Palaiseau, France,
\\
\email{denis.bernard@in2p3.fr}
and
\\
Stanley D. Hunter \at NASA/Goddard Space Flight Center, Greenbelt Road, Greenbelt, MD 20771, USA
\\
\email{stanley.d.hunter@nasa.gov}
 and
 \\
Toru Tanimori \at Graduate School of Science Kyoto University, Kyoto, Japan 606-8502,
\\
\email{tanimori.thoru.6ekyoto-u.ac.jp}
}

%
\maketitle

~

\abstract{The detection of photons with energies greater than a few tenths of an MeV, 
interacting via Compton scattering and/or pair production, faces a number of difficulties. 
The reconstruction
of single-scatter Compton events can only determine the direction of the incoming
photon to a cone, or an arc thereof and the angular resolution of pair-conversion
telescopes is badly degraded at low energies. 
Both of these difficulties are partially overcome if the density of the interaction medium is low. 
 Also no precise polarization measurement on a cosmic source has been
 obtained in that energy range to date.
 We present the potential of low-density high-precision homogeneous
 active targets, such as time-projection chambers (TPC) to provide an
 unambiguous photon direction measurement for Compton events, an
 angular resolution down to the kinematic limit for pair events, and
 the polarimetry of linearly polarized radiation.
}

\section{Keywords} 

Gas time-projection chamber (TPC),
Compton scattering,
pair conversion,
point spread function,
micropattern gas detector (MPGD),
electron-tracking Compton camera (ETCC),
polarization
\section{Introduction}
\label{sec:intro}

Gamma-ray astronomy proceeds by the analysis of the conversion of
individual photons with an atom of a telescope, either
\begin{itemize}
\item
the Compton scattering on an electron, $\gamma e^- \to \gamma e^- $,
in the approximate energy range
$1 \,\kilo\electronvolt < E < 100 \,\mega\electronvolt$, or

\item
 by the creation of an $e^+e^-$ pair
 ($\gamma Z \to e^+e^-Z$ ``nuclear'' conversion, above a threshold $E > 2mc^2$; 
 $\gamma e^- \to e^+e^-e^-$ ``triplet'' conversion, $E > 4mc^2$), where $m$ is the electron mass.
\end{itemize}

At lower energies, X-rays are detected via their photo-electric
interaction with matter, see Volume 1 of this Handbook.

A number of detector techniques have been developed (see Chapter
Telescope concepts in gamma-ray astronomy in Vol. 2, Sect. IV), among
which some use an homogeneous active target, that is, a device in
which the photon converts and the trajectory of the charged particle(s)
in the final state (electron and/or positron) are measured, that
consists of an homogeneous medium.
Among these,
a TPC (time projection chamber) \citep{Marx:1978zz} is a device in
which a chunk of matter is immersed in an electric field;
 charged particles ionize the atoms/molecules on their way
through the detector, after which the produced electrons and positive
ions drift in the electric field towards an anode and a cathode,
respectively, where their arrival creates an electric signal that can
be measured.
The collecting anode can be segmented in a two-dimensional (2D) series
of pads, or in a two-fold set of strips, something which provides a 2D
image of the flow of electrons that are ``falling'' on the anode at a
particular time.
The measurement of the drift duration provides the third coordinate:
the collection and the analysis of these images as a function of time
provides a full, 3D, picture of the charged particles in the final
state for the particular event of the conversion of the photon.

For gases, 
the amount of charge produced by ionization is small, typically tens
to hundreds of electrons per $(\centi\meter \cdot \bbar)$, so an
amplification in the gas is needed, that was performed with multi-wire
proportional chambers (MWPC),
that have later been replaced by micropattern gas detectors (MPGD)
such as micromegas \citep{Giomataris:1995fq}, gas electron multiplier
(GEM) \citep{Sauli:1997qp} or micro-pixels \citep{Ochi:2000js}.

For two-track final states, 
in the case of a two-fold series of strips, reconstructing the full 3D
image of the event requires solving a two-fold ambiguity (i.e. the
matching of either of two tracks in the ($x, t$) plane to either of two
tracks in the ($y, t$) plane) something that is made easy
(Fig. 11 of \citep{Bernard:2014kwa})
by the violent
variation of the energy deposited along the track, the distribution of
which presents a large tail at high values (the Landau distribution).
The ambiguity can also be solved with a three-fold series of strips,
for example at 120\degree of each-other, in a configuration known as
$(x,u,v)$.

In contrast with MWPCs for which the wire pitch is limited to values
larger than a couple of millimeters, the collecting structure of MPGDs
can be much smaller, enabling spatial resolutions down to
$50\,\micro\meter$ \citep{Thers:2001qs}.

In addition to measuring precisely the geometry of the conversion, 
TPCs can also provide precise measurements of the energy deposition in
the gas, something which is key to high-precision Compton scattering,
and of the energy deposition per unit length, 
$\dd E / \dd x$, 
which enables the rejection of $z>1$ cosmic-ray ions, as 
$\dd E / \dd x$ is proportional to $z^2$ (Eq. (34.5) of 
\citep{Zyla:2020zbs}).
The TPC concept proves to be particularly powerful for the polarimetry
of the incoming radiation, see
Chapter: Time projection for polarimetry in Sec. II of Vol. 1 (X-rays)
and
Chapter: Design of gamma-ray polarimeters in Sec. IV of Vol. 2.
Reviews on TPCs can be found in Sec. 35.6.5 (gas TPCs) and in
Sec. 36.4 (liquid TPCs) of \citep{Zyla:2020zbs} and in
\citep{Hilke:2010zz,Fujii:2014}.
The transport of electric charges in matter, drift detectors and
their use is described in \citep{Sauli:1977mt,Blum:Rolandi,Sauli:2014},
the simulation of gas detectors can be performed by the Garfield++
software package \citep{Veenhof:1998tt} and the RD51 Collaboration
\citep{RD51} is coordinating the efforts of the community to develop 
gas detector technologies.
Various configurations have been developed to meet various needs, such
as
``cartesian'' TPCs (uniform electric field, e.g. \citep{Atwood:1991bp}),
``radial'' TPCs (e.g. \citep{Obertelli:2014uaa}) and
``spherical'' TPCs (e.g. \citep{Giomataris:2008ap}).

TPCs are mainly used as inexpensive (in terms of the number of electronics
channels) hyper-high-granularity trackers for low (Hz to kHz) rate
experiments, due to the duration of the drift (microseconds for the
electrons to milliseconds for the ions), as most users wish to avoid
event pile-up.
The TPC itself, though, can cope easily with much higher rates (e.g. a
$\overline p$ annihilation rate of 20\,MHz 
\citep{Fabbietti:2010fv,Ball:2012xh}), as long as the electric charge
of the ion back-flow does not disturb the electric field, in which
case gating the amplification device is advised \citep{Nemethy:1982uw}.
The complexity load is then transferred to the event-reconstruction
software (See Fig. 13 of \citep{Rauch:2012hp}).
It should be noted that most of the positive ions are created in the
amplification volume of the MPGD and that in the case of the
micromegas and of the GEM, a large fraction of them is collected by
the grid, and therefore does not escape to the drift volume, something
that limits the induced nuisance.

An important limitation of TPCs in their use for high-energy-physics (HEP) 
is their reliance on the information provided by other sub-detectors
to build a trigger, and in particular to define the start time of the
drift.
Difficulties in obtaining a sufficiently large background noise
rejection factor can induce Data Acquisition (Daq) dead time, and lead
to a severe reduced trigger efficiency, as observed for example during
the balloon test flight of the Liquid Xenon Gamma-Ray Imaging
Telescope (LXeGRIT) TPC prototype \citep{Curioni:2007rb}.
For a stand-alone TPC in space, a trigger-less, continuous, autonomous
mode \citep{Rauch:2012hp,Hunter:2014} can be considered.

Dielectric gases that allow the free drift of injected electric
charges, such as noble gases or alkanes, can be used, that have
negligible electron attenuation over drift lengths of several meters.
As electro-negative impurities in the gas can trap drifting electrons,
maintaining a good gas purity on the long term is important.
For example, the degradation of the gas in the EGRET spark chambers
required regular gas changes and caused the instrument to be operated
only intermittently in the later part of the Compton Gamma Ray
Observatory mission \citep{EGRET:1999}.
Actually the dose rate of ionizing radiation that is inducing the
production of these deleterious chemical species is much lower in
orbit than in detectors on HEP experiments, and spark chambers
are well known to be very powerful quencher polymerizing devices.
Using the HARPO TPC prototype, that includes a number of elements that
are potentially harmful to the gas purity (PVC, epoxy ..) continuous,
perfectly stable operation was demonstrated for over six months just
filtering out the oxygen contamination \citep{Frotin:2015mir}.
The test of a TPC prototype designed using high-vacuum technology for
the Gravity and Extreme Magnetism Small Explorer (GEMS) X-ray
polarimeter Instrument (XPI) demonstrated a gas-fill estimated
lifetime larger than 23 years \citep{Hill:2013};
they estimated the integrated charge expected on a space mission to be
of the order $10^{-3}\milli\coulomb/\milli\meter^2$ \citep{Hill:2013}.
This can be compared to the performance of micromegas and GEM devices
for LHC detectors, that can stand $> 20 \milli\coulomb/\milli\meter^2$
(which corresponds to $> 6.\,10^{11} \MIP/\milli\meter^2$) without
degradation \citep{Titov:2004rt}, though with flushing some amount of
fresh gas.
A gas purifier system for the close-loop operation of Resistive Plate
Chambers (RPC) at the LHC, with high recirculation fractions, has
shown stable operation up to an integrated charge of
$0.5 \milli\coulomb/\milli\meter^2$ \citep{Capeans:2013uma}).
Alkanes can be replaced by CO$_2$ to alleviate the polymerization
issue.
In that case, at extreme irradiation doses, the next contaminant of
concern is found to be silicone, in particular from pump-oil
outgassing (see, e.g., \citep{Zimmermann:2004gta}).
Despite the huge body of knowledge accumulated on ageing of detectors
exposed to extreme irradiation doses (but with some amount of fresh
gas flushing), or sealed detectors in operation for an extended
duration (but without an irradiation dose rate commensurate with that
on a space mission), it is fair to state that a smooth decade-long
operation of a sealed (possibly with recirculation and filtering)
proportional gas detector exposed to an appropriate level of ionizing
radiation is still to be demonstrated.

MPGDs are prone to sparking, especially when exposed to a
charged-particle beam, with a rate that increases with the applied
amplification voltage.
The limiting mechanism was understood after it was observed that the
spark production rate is much larger from a pion beam than from a muon
beam of similar energy and intensity \citep{Thers:2001qs}:
hadrons can undergo hadronic interactions with the atoms of the
detectors, after which a nuclear fragment can deposit a huge amount of
energy in a very small volume at the end of its trajectory, due to the
$1/\beta^2$ variation of $\dd E / \dd x$ (the Bragg peak)
\citep{Zyla:2020zbs}.
The phenomenon is particularly prevalent for micromegas, for which the
transverse size of the avalanche is small, of the order of
$5 - 10\,\micro\meter$, and/or when a magnetic field is applied, that
clamps the transverse diffusion of the drifting electrons.
The problem is now routinely solved using an additional resistive layer
\citep{Alexopoulos:2011zz}.

\section{Charged particles production and transport in a medium}
\label{sec:transport}

\subsection{Ionization}
\label{subsec:ionization}

The average energy lost by a charged particle traversing a piece of
matter is described by the Bethe equation (eq. (33.5) of
\citep{Zyla:2020zbs}), that shows a sharp rise proportional to
$1/\beta^2$ at low energies, a minimum at about $\gamma\beta =3$ and a
slow rise at higher energies ($\beta$ is the particle velocity
normalized to that of light and $\gamma$ is its Lorentz factor).
The specific loss rate at the minimum (MIP, minimum ionizing particle)
is proportional to $(z^2 Z)/(A \beta^2)$ and is therefore
approximately independent of the nature of the gas, at
$(1/\rho)\dd E / \dd x \approx 2 \,\mega\electronvolt \gram^{-1} \centi\meter^2$,
where $\rho$, $Z$ and $A$ are the density, the atomic
number and the mass number of the target 
(Tab. 6.1 of \citep{Zyla:2020zbs}),
and $z$ is the electric charge of the particle normalized to the charge of 
an electron.
Part of the interactions results in the ionization of the atoms,
with an effective energy lost per produced electron of 20 to 40\,eV
(Tab. 1 of \citep{Sauli:1977mt}).

\subsection{Drift, diffusion}
\label{subsec:drift:diffusion}

As the ionization electrons drift in the electric field of the TPC,
they undergo elastic collisions with the atoms and molecules of the
gas, a process by which they lose some of their kinetic energy and by
which they are deflected.
Between two collisions, the electrons are accelerated in the field and
follow a ballistic trajectory.

At low electric fields, the resulting average drift velocity, $v$, is
proportional to the electric field, ${\cal E}$, with typical values of
the mobility $\mu \equiv v / {\cal E}$ of 
$\mu p = (0.5 - 2) \, \centi\meter^2 \, \bbar \, \volt^{-1} \, \second^{-1}$
for most positive ions (Tab. 4 of \citep{Sauli:1977mt}), and
$\mu p = (0.5 - 2) \, 10^4 \, \centi\meter^2 \, \bbar \, \volt^{-1} \, \second^{-1}$
for electrons, where $p$ is the gas pressure.
On average, the deflections result in an isotropic,
Gaussian-distributed spread of the distribution of the electron
position that is named diffusion, with an RMS along any of the three
directions of space
$\sigma = \sqrt{2 D t}$ with $L = v t$, where $t$ and $L$ are the
duration and the length of the drift and $D$ is the diffusion
coefficient
\citep{Sauli:1977mt}:
\begin{equation}
\sigma = \sqrt{\gfrac{2 kT L}{e {\cal E}}}.
\end{equation}

The spread can also be expressed as 
$\sigma = \sqrt{d L}$ where
$d = {2 kT }/{(e {\cal E})}$: 
the low-field RMS spread is independent on the
material used: at a given temperature, the only way to decrease 
$\sigma $ is to increase the electric field.

At high electric fields, the rise of the electron drift velocity with
the electric field saturates, the electron velocity distribution is
characterized by a temperature that is larger than the ambient, and
the diffusion process becomes anisotropic, with a diffusion
coefficient in the transverse plane $D_T$ and in the longitudinal
direction $D_L$ that differ increasingly at higher field.
In pure argon, for example, the spread would amount to
$0.11 \,\centi\meter / \sqrt{ \centi\meter }$ at
$ {\cal E} / p = 1 \,\kilo\volt/ (\centi\meter \cdot \bbar)$
while the thermal limit at 300\,K at that field is
$80 \,\micro\meter / \sqrt{ \centi\meter }$.

Therefore a (usually small) fraction of multi-atomic (means here $>$ 2
atoms) molecular gas is added.
These molecules have vibration and rotation degrees of freedom that
enlarge the cross section of inelastic interactions from fast
electrons, something which ``cools'' the electron bunch, and also they
have absorption bands that mitigate the potentially catastrophic
discharges induced by photo-electric effect on the detector cathode by
impacts of avalanche-created U.V. photons -- hence their appellation
of ``quenchers''.

Most gas TPCs use a mixture based on a large fraction of a noble
gas together with a multi-atom molecular quencher
(example an alkane). 
These ``fast gases'' allow at the same time a high electron-drift
velocity of several $\centi\meter/\micro\second$ and a small diffusion
coefficient.
For argon-isobutane mixtures, for example, the transverse 
coefficient $d_T$ plateaus above
$ {\cal E} / p \approx 50\,\volt/(\centi\meter \cdot \bbar)$
at a value that is proportional to 
$1/\sqrt{p_r}$, where $p_r$ is the quencher partial pressure
(e.g. $170\, \micro\meter/\sqrt{ \centi\meter}$ for a 4-bar 90:10 mix).
In these conditions, the longitudinal coefficient $d_L$ is a smoothly decreasing
function of $ {\cal E}$, with little dependence on the quencher
fraction and gas pressure, and amounts to
$180\, \micro\meter/\sqrt{ \centi\meter}$ for 4-bar 90:10 at
$ {\cal E} = 500 \,\volt/\centi\meter$).
It should be noted that some amount of diffusion, with a spread
commensurate with the pitch size of the signal collection
segmentation, enables the optimization of the spatial resolution of
individual measurements (Fig. 7 of \citep{Arogancia:2007pt}).
Charge amplification can be easily performed in these gases with gains
of several $10^4$ \citep{Attie:2009zz,Veenhof:2010} in current MPGD
structures such as a multi-layer GEM \citep{Sauli:1997qp} or a
micromegas \citep{Giomataris:1995fq}.

\subsection{Negative Ion Technique}
\label{subsec:negative:ion}
 
Another option to reduce the diffusion of a TPC is to add a mildly
electronegative component to the TPC gas. This is referred to as the
negative ion (NI) technique, developed for TPC dark matter searches,
which originally used carbon disulfide (CS$_2$)
\citep{Martoff:2000}
 \citep{Martoff:2005}.
Subsequently, nitromethane (CH$_3$NO$_2$) was also used
\citep{Martoff:2009}.
The NI technique has been used in several different applications
\citep{Snowden-Ifft:2003}
 \citep{Son:2010}.

In the gas of a negative ion TPC (NI TPC), the NI component scavenges
the free, ionization, electrons, within $\approx 100\,\micro\meter$ of
their point of origin, to form negative ions, which 
then drift in the TPC gas.
Compared to the electrons, which drift at super-thermal speed, the
negative ions, being much more massive, remain in thermal equilibrium
with the gas molecules and drift at the thermal velocity of the
gas. Consequently, the diffusion of the NIs is reduced to the thermal 
diffusion limit, see Figure 6 in
\citep{Dion:2011}.

The NI technique reduces the diffusion coefficient to the thermal
limit of $d = 80\,\micro\meter/\sqrt{\centi\meter}$, but at the cost of
a much lower drift velocity $v = 2 \,\centi\meter / \milli\second$
\citep{Hunter:2018wns}.
This velocity is three to four orders of magnitude smaller than for
electrons in a typical fast TPC gas \citep{Peisert:1984}.
This slow velocity can make the detector more susceptible to
background pile up.
However, the high granularity of a TPC and the reduced diffusion and
slow drift velocity allow image processing techniques to be employed
to differentiate the high-energy photons interacting to produce
electron pairs from the Compton low-energy recoil electrons produced
by sub-MeV photons and cosmic ray tracks.
Software techniques have been explored to address this problem
\citep{Garnett:2020}.
Application of the negative ion technique to the Advanced Energetic
Pair Telescope, and software solutions to deal with the slow drift
velocity, is discussed in Subsection \ref{subsec:pair:achievements:prospects}.

\subsection{Energy measurements}
\label{subsec:energy:measurements}
 
Gas TPCs are thin detectors from which high-energy electrons escape
(1\,MeV electrons have an $0.5 - 0.7\,\gram/\centi\meter^2$ range in
most gases \citep{NIST:electrons}, e.g. 3.6\,m in 1-bar argon).
Also, the measurement of the track momentum by the analysis of the
multiple deflections induced by multiple scattering in the detector,
which is a powerful method for denser detectors, is too imprecise for gas TPCs 
\citep{Frosini:2017ftq}.
So it is difficult to measure the energy of the leptons, and therefore
of the candidate photon, from the TPC alone.

Either the TPC can be immersed in a magnetic field, to become a
high-precision magnetic spectrometer, or the TPC can be complemented
with an additional detector that either measures the momentum of the
track(s),
such as a transition-radiation detector (TRD) 
\citep{Wakely:2004gg}
or the total energy of the final state, such as a calorimeter.

\subsection{Magnetic field}
\label{subsec:magnetic:field}
 
Most gamma-ray telescopes on space missions do not include a magnetic
spectrometer.
The AMS experiment on the ISS does \citep{Bourquin:2005ef}, however,
and has a sensitivity to gamma rays \citep{Beischer:2020rts}.
The presence of a magnetic field has important consequences on the
performances of the detector;

\begin{itemize}
\item In the first place, the curvature enables a measurement of 
 track momenta;
 
\item Also, low-momentum electrons are clamped to spiral and deposit
 their energy in a small volume, something that can help the trigger
 system survive the large Compton-induced single-track background;

\item Spiraling electrons might be an issue though, for the correct
 tracking of low-energy electrons in a Compton camera;

\item The transverse diffusion of the drifting electrons is greatly
 reduced, as the ballistic flight between collisions is along
 spiraling helices instead of parabolas.

\item Last but not least, at very high electron momenta where multiple
 scattering can be neglected, in the presence of a non-zero magnetic
 field, $B$, the track fit includes a track curvature while for $B=0$
 a straight line is fitted.
The non-zero correlation between the curvature and the angle of the
track at vertex induces a degradation of the single-track angle
resolution of a factor of 4 with respect to the fit that assumes $B=0$
(compare, for example, eqs. (5) and (9) of \citep{Regler:2008zza}).
\end{itemize}

\subsection{Absolute time measurement}
\label{subsec:time:measurement}
 
The ability to measure precisely the absolute time of a gamma-ray
conversion can be of interest, for example for pulsars or for
transients.
MPGD are fast devices, that enable event-time measurements with a precision
of a couple of nanoseconds with a suitable electronics for tracks
that are crossing them.
This could allow building a trigger from a telescope consisting of a 3D set
of individual TPC modules \citep{Bernard:2014kwa}, by forming
multi-module coincidences.

For events for which the track(s) exit(s) by the side of the TPC
without crossing the amplification structure, the degenerescence
between the start time and the vertical position of the event in the
TPC fundamental relation,
$z = v (t -t_0)$, 
makes it difficult to determine separately the vertex vertical
position and the conversion time.
In these cases, the variation with drift time of the track width due
to diffusion-induced spread enables a measurement
\citep{Antochi:2020hfw}, though with a much degraded resolution.

\section{Electron-Tracking Compton Camera with Gaseous Time-Projection Chamber}
\label{sec:etcc}

The simplest Compton telescopes are based on the analysis of
one-scatter events, $\gamma e^{-} \to \gamma e^{-}$, after which the
scattered photon propagates in the detector until it is absorbed: the
position and the energy of the scattered electron and photon are
measured, and the properties of the incident photon are inferred from
this information.
With that method, the direction of the incident photon is found to lie
on a cone, something which makes the analysis of an extended field of
view containing several sources complicated.

\begin{figure}[htbp]
 \begin{center}
 \includegraphics[width=3.5in, trim=4 4 4 4, clip=true]{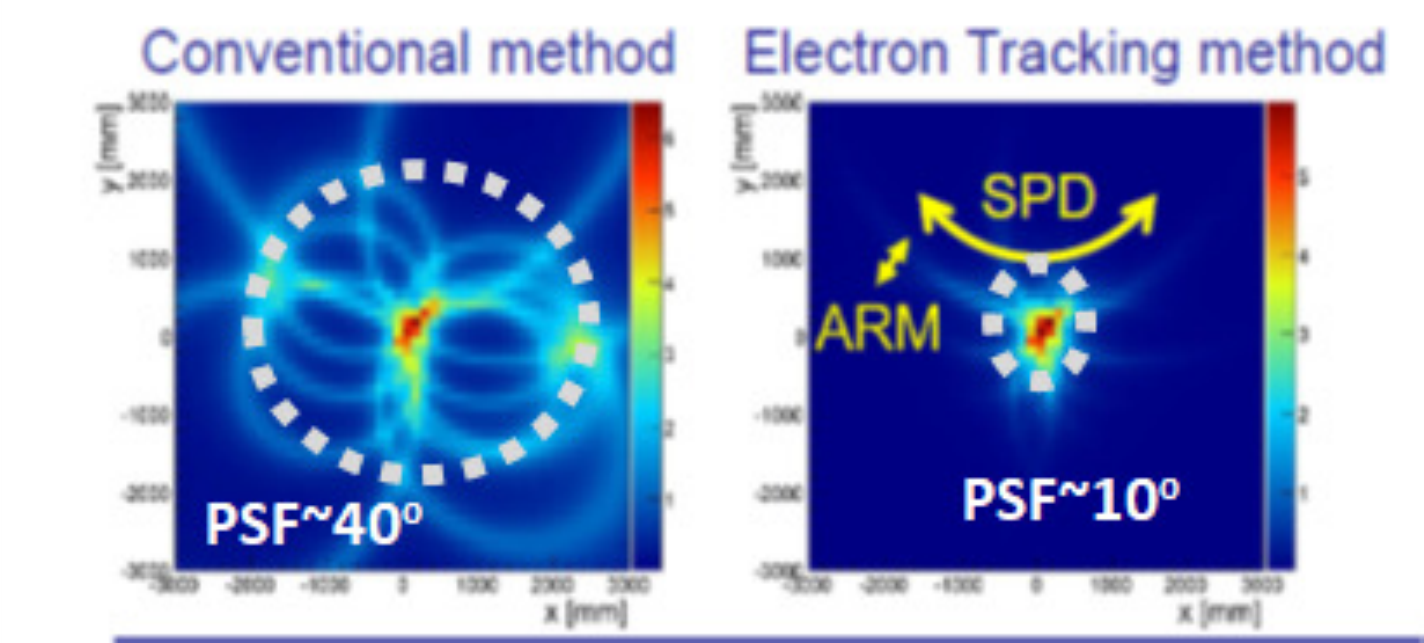}
 \end{center}
 \caption{Images of a point source with a Compton Camera (CC, left) and with an
  electron-tracking Compton camera (ETCC, right).
  Adapted from \citep{Tanimori2020}.
 \label{fig1}}
\end{figure}

In more elaborate telescopes, the direction of propagation of the
recoiling electron is tracked, so that the cone is reduced to an arc,
something that simplifies the image analysis
(Fig.~\ref{fig1} left).

The present Section addresses the extreme case for which the tracking
of the electron is so precise that the analyst can use a bijection
between the measured observables for a given Compton scatter event and
the direction of the incoming photon
(one-to-one mapping, Fig.~\ref{fig1} right),
in a scheme that is named ETCC
 (electron-tracking Compton camera)
\citep{Tanimori2004,Tanimori2015,Tanimori2017}.
We will see that a low-density high-precision active target, such as a
gas TPC, is well suited for that purpose.

\subsection{How to realize complete bijection imaging for MeV gamma rays}
\label{subsec:bijection}

\begin{figure}[htbp]
 \begin{center}
\includegraphics[width=3.in]{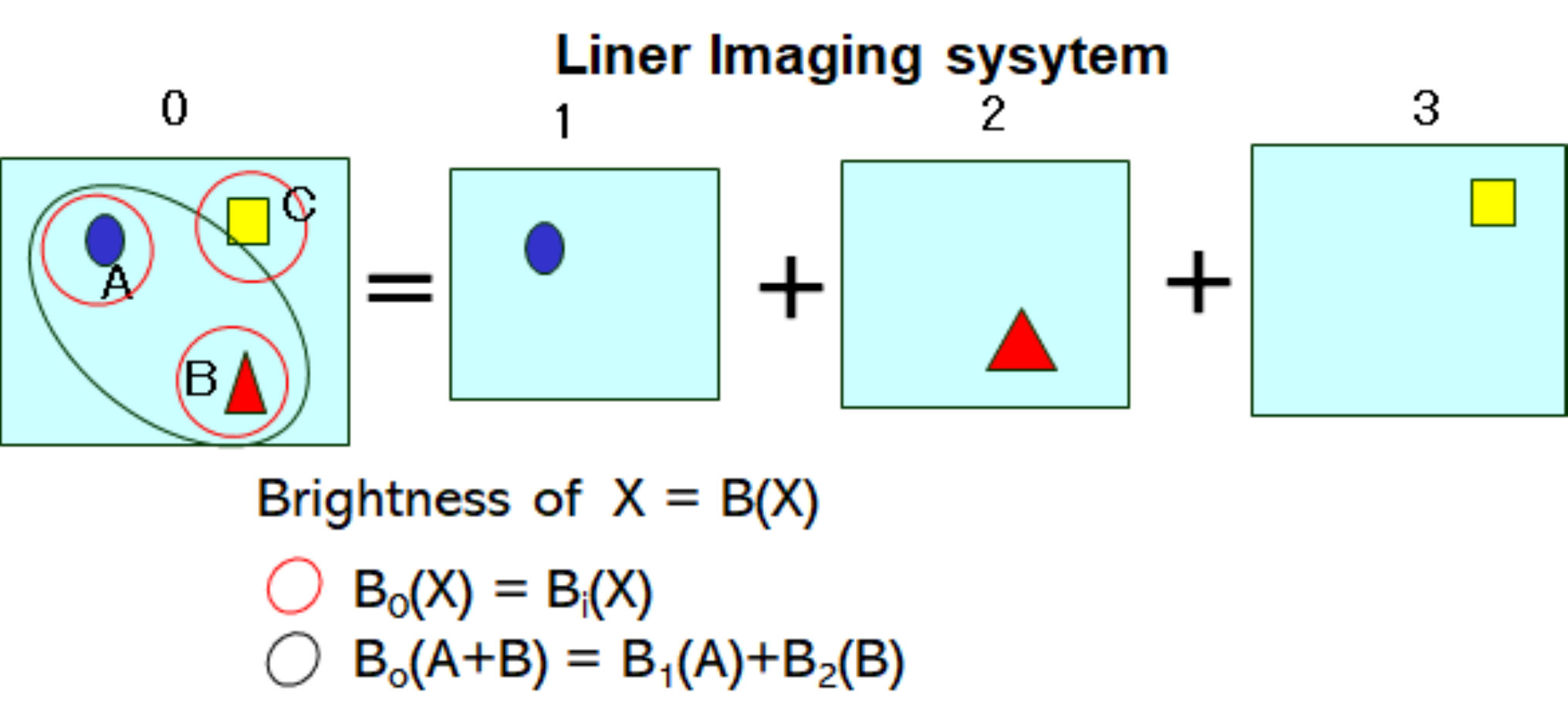}
 \end{center}
 \caption{Explanation of the linear imaging system, which is essential for quantitative imaging analysis.
 }
 \label{fig2}
\end{figure}
\begin{figure}[htbp]
 \begin{center}
\includegraphics[width=3.in]{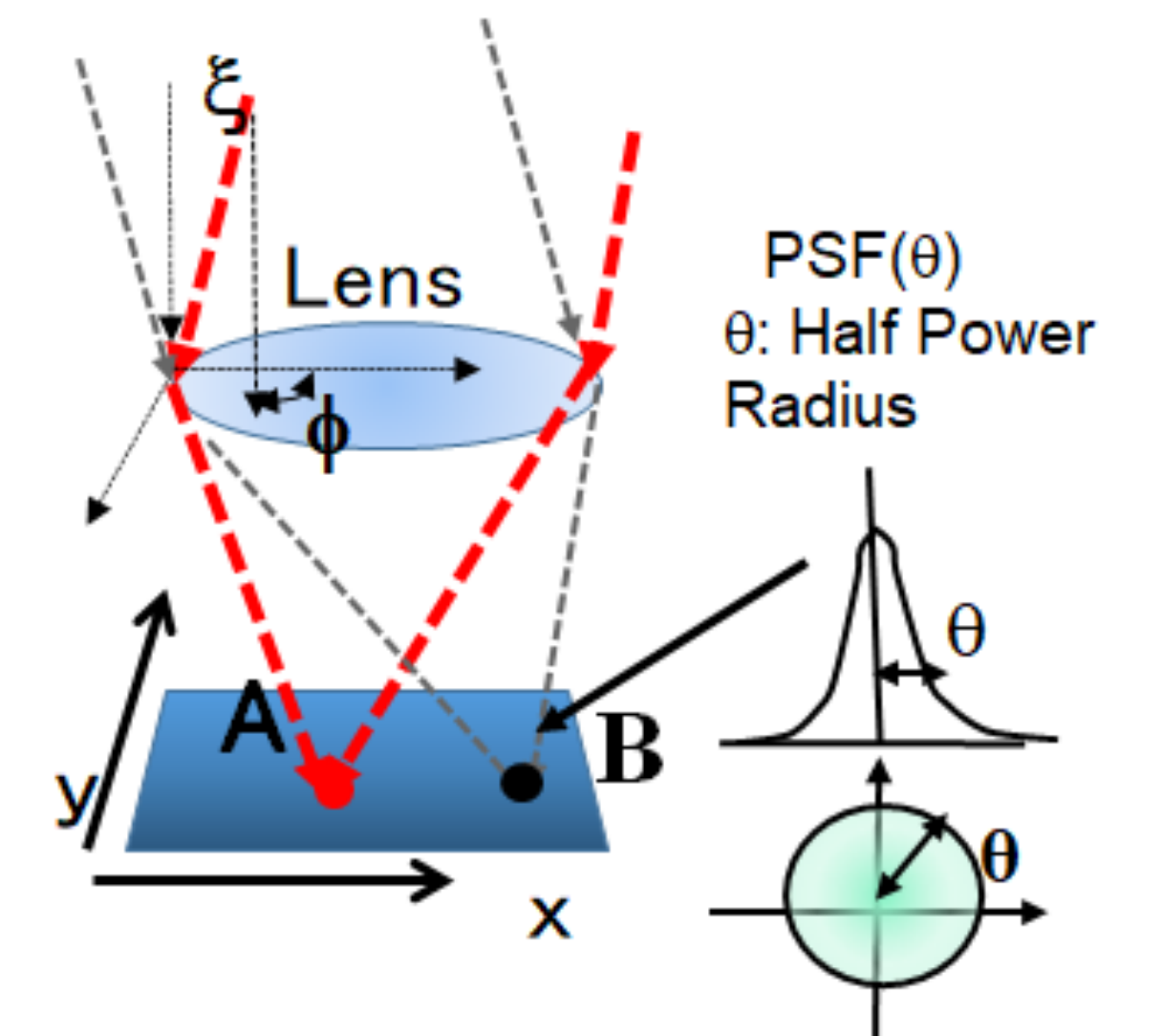}
 \end{center}
 \caption{Schematic of the bijection image and PSF.
 }
 \label{fig3}
\end{figure}

In astronomical imaging, the intensity of each point in the image is
independent, and linearity is preserved, as shown in Fig.~\ref{fig2}.
This is the minimum requirement for maintaining quantitativeness in
image analysis in general.
Quantitative imaging is an essential technology, e.g., in radio and
X-rays, because the electromagnetic waves can generally be refracted
and reflected, as well as simply focused according to optical
principles using a reflector and a lens.
As shown in Fig.~\ref{fig3}, reflectors and lenses make a bijection to
each point on the image surface while preserving the intensity of rays
in each direction at a size larger than the point spread function
(PSF).
Therefore, the intensity of each point separated from the PSF of the
image is guaranteed to be linear, and the intensity can be accurately
measured.
The PSF is defined as the minimum angular distance that guarantees
linearity.
Here, we use a half-power angular radius (degrees) as the PSF.
The linear optical system is shown in Fig.~\ref{fig2}.

\begin{figure}[htbp]
 \begin{center}
\includegraphics[width=3.54444in]{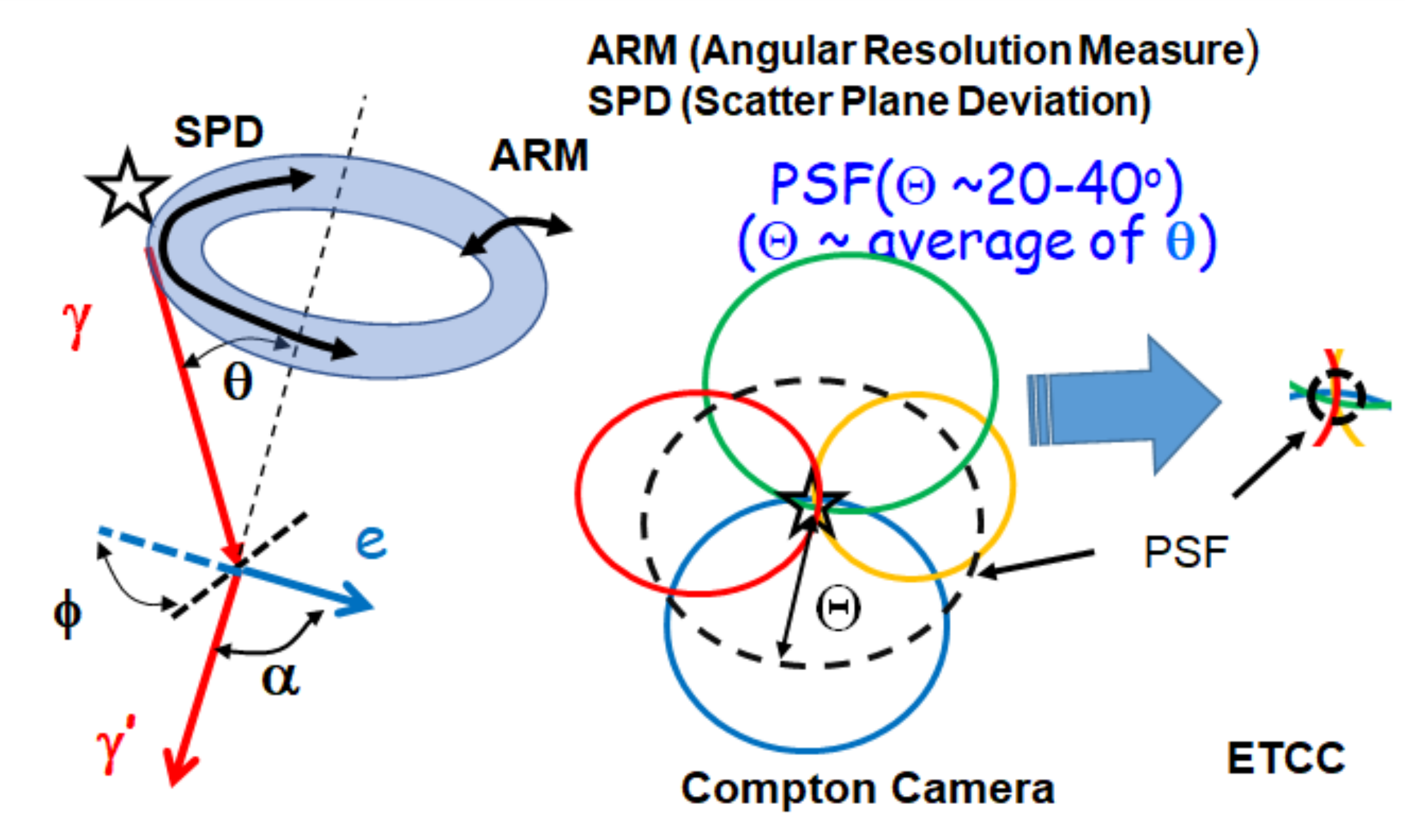}
 \end{center}
 \caption{Schematic of the Compton scattering kinematics. The
 parameters $\theta$ and $\varphi$ represent the Compton-scattering
 angle and electron-azimuthal angles of the incident gamma ray,
 respectively, on the Compton coordinates. (right) Schematic
 explanation of the PSF($\Theta$) of the CC and ETCC. See the text
 for the definition of $\Theta$. }
 \label{fig4}
\end{figure}
\begin{figure}[htbp]
 \begin{center}
\includegraphics[width=3.02778in]{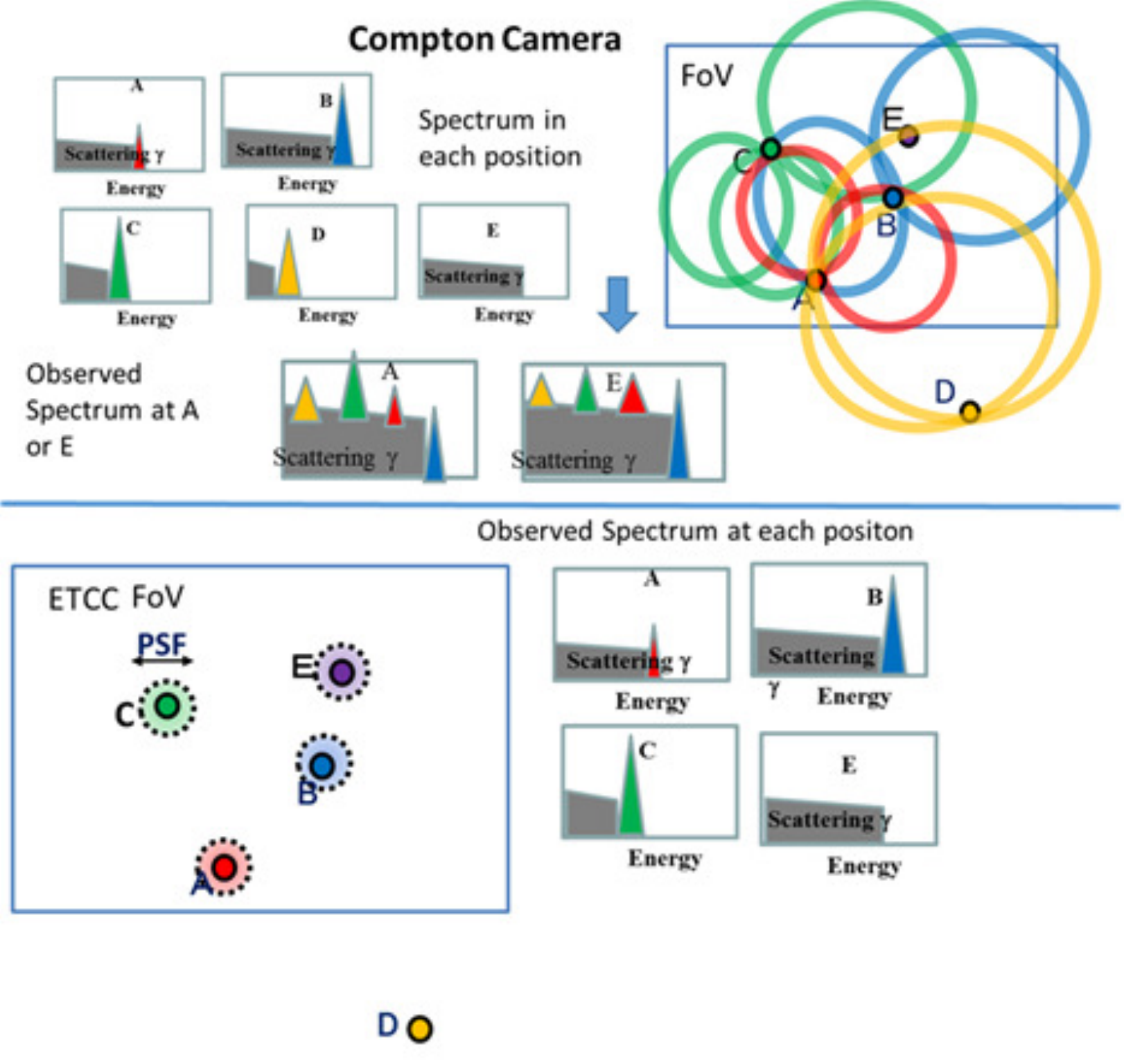}
 \end{center}
 \caption{Schematic of the nonlinear system of the CC and the linear
 one of the ETCC. The observed spectra for each target point
 clearly indicate that the CC cannot perform imaging spectroscopy.
 }
 \label{fig5}
\end{figure}

However, gamma rays have strong quantum properties, and
refraction/reflection cannot be used.
Therefore, the only way to determine the direction of the incident
gamma ray is to solve the kinematics of Compton scattering for each
gamma ray.
This was attempted using a CC, starting in the 1970s.
However, because the CC solves the equation of motion of Compton
scattering incompletely owing to the lack of the direction of the
recoil electron, only the elevation angle in the arrival direction can
be obtained, as shown in Fig.~\ref{fig4}.
Therefore, the gamma-ray direction is given only as an annulus.
The gamma-ray distribution is estimated by superimposing this annulus
on the image.
The annulus spreads over the entire field of view (FoV) to several
tens of degrees or more, making it difficult to define the PSF, and
the information at each point on the image is strongly mixed and
influenced, as shown in Fig.~\ref{fig5}, where bijection cannot be
satisfied by a CC.
Thus, the CC lacks nearly half the information and loses
quantification, which is the essence of image analysis.
Sometimes, the CC employs the ``effective PSF'' using only the angular
resolution of the zenith angle obtained by the CC, which is identical
to the proper two-dimensional (2D) PSF \citep{Schonfelder1993}.
However, gamma-rays within the ``effective PSF'' are surely and
strongly influenced by the large flux of gamma-rays outside of it and
are never separated from background gamma-rays coming outside the
effective PSF, whereas telescopes for other wave bands having a proper
PSF easily separate gamma-rays within and outside the PSF.
Thus, the CC is certainly a nonlinear optical point system, but all
other telescopes, including the ETCC, are linear systems.
In particular, considering the large radiation background from the
satellite and instruments due to collision with cosmic rays, as
mentioned in the next Section, a rigid quantitative imaging system is
needed to remove this background.

\begin{figure}[htbp]
 \begin{center}
\includegraphics[width=3.83333in]{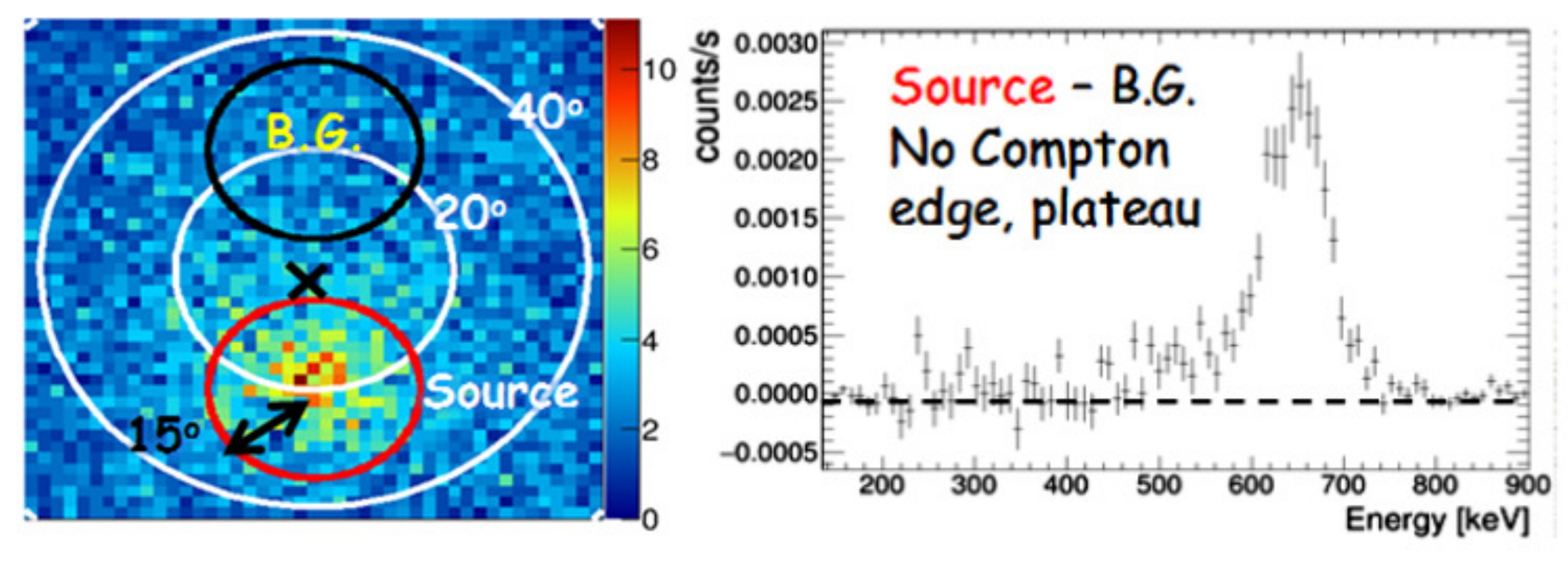}
 \end{center}
 \caption{Image of the \textsuperscript{137}Cs source at the
 off-axis with a polar angle of 20$\degree$ measured by SMILE-II. Red
 and black circles, each having a radius of 15$\degree$, indicate
 source and background regions, respectively. (right) The spectrum
 after air-scattered gamma-rays within the PSF are removed, by using
 the region symmetric about the center of the FoV (black circle) as
 the background region. Adapted from \citep{Tanimori2017}. }
 \label{fig6}
\end{figure}

To overcome this problem, it is necessary to completely solve the
kinematics of Compton scattering, which can be achieved by measuring
the direction of the recoil electron.
As shown in Fig.~\ref{fig1}, the ETCC focuses MeV gamma rays emitted
from the point source to one point, similar to optical telescopes.
Only the determination of both the angles of incident gamma-rays surely
enables selection of the gamma-rays in the FoV from background gamma-rays coming
from the outside with the resolution of the PSF and to estimate the
leakage of the background from the outside of the FoV, as shown in
Fig.~\ref{fig5} \citep{Tanimori2017, Tanimori2015}.
Additionally, such a proper PSF gives us the ON region, as well as the
interested region and OFF-regions, which are hardly affected by the ON
region in the same FoV.

The background-subtracted signal is obtained using the ON-OFF method,
as shown in Fig.~\ref{fig6} \citep{Tanimori2015}.
Thus, the precise definitions of the FoV and background regions make
it possible to quantitatively evaluate the detection of the signal.
These methods and concepts are essential and common in all fields of
astronomy and science using imaging analysis.
Furthermore, the intensity independence of each point above the PSF in
the image ensures that the spectrum of each point comes from that
location, as shown schematically in Fig.~\ref{fig5}, which is called
imaging spectroscopy \citep{Tanimori2015}.
As explained previously, most universal imaging analysis methods are
based on a clear definition of the PSF, and a sharper PSF provides
better quantities from imaging analyses in general.

\subsection{Background rejection in ETCC}
\label{subsec:background:rejection}
 
\begin{figure}[htbp]
 \begin{center}
\includegraphics[width=4.in]{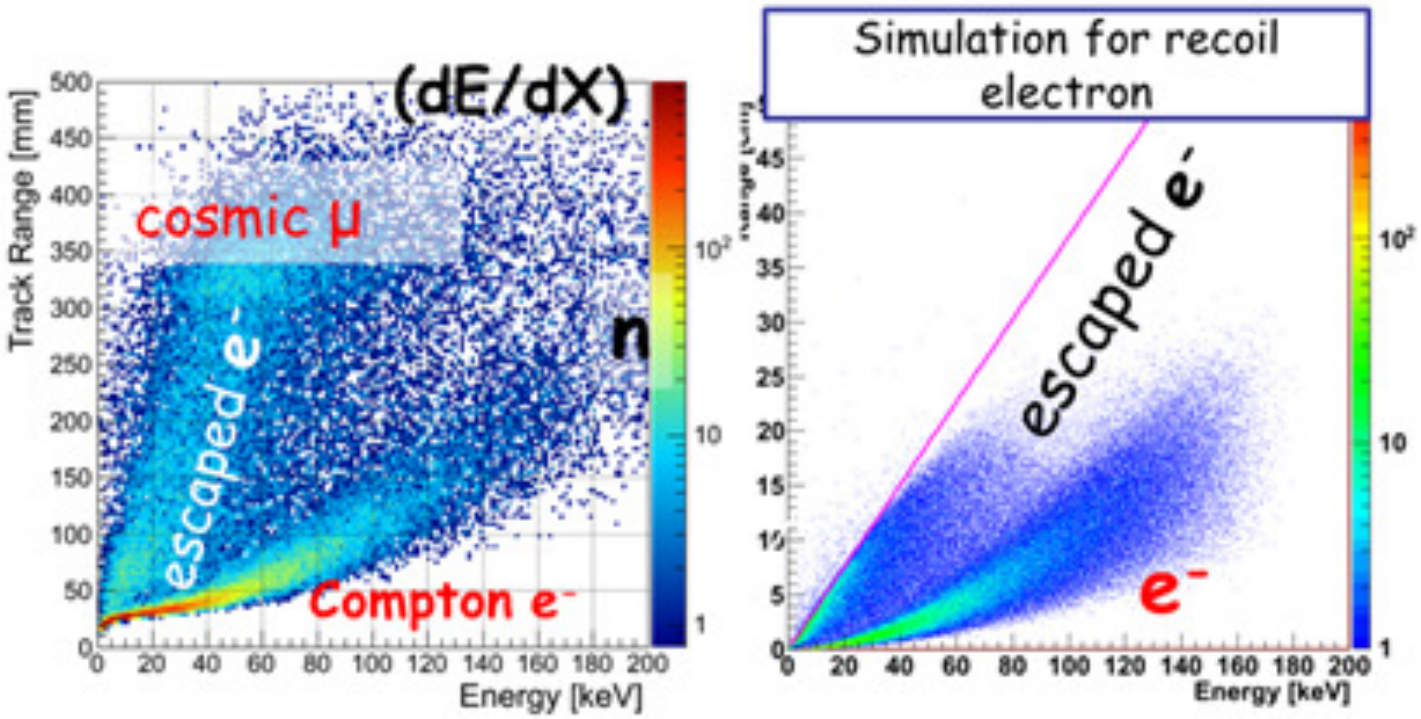}
 \end{center}
 \caption{Observed and simulated scatter plots of track energy and
 its length in SMILE-II. The ratio is $d$E/$d$x, and Compton recoil
 electron is clearly identified from background events.
 Adapted from \citep{Tanimori2015}. }
 \label{fig7}
\end{figure}
\begin{figure}[htbp]
 \begin{center}
\includegraphics[width=2.5in]{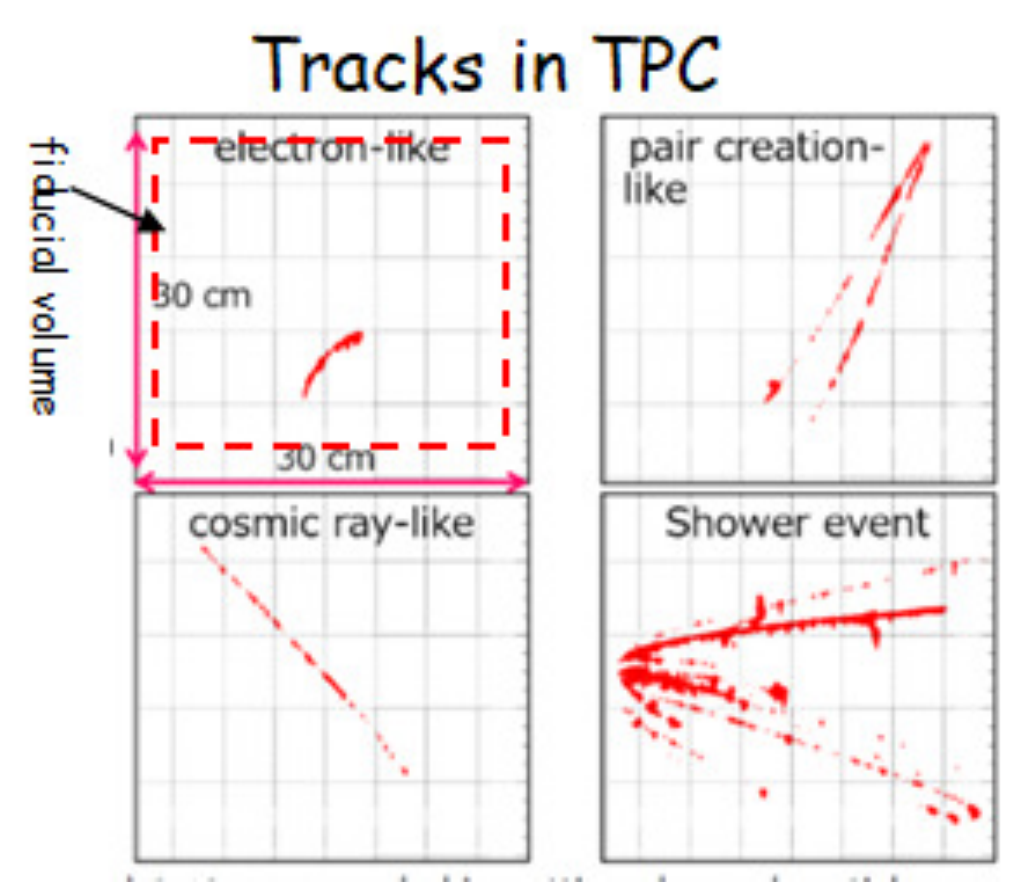}
 \end{center}
 \caption{Event topologies observed by the SMILE-2+ TPC during the balloon ascent in 2018. Adapted from \citep{Tanimori2020}.
 }
 \label{fig8}
\end{figure}

Another serious problem is the large radiation background in space.
Although gamma backgrounds coming outside the FoV are significantly
reduced by the linear imaging system, other types of backgrounds, such
as neutrons, missing charged cosmic rays, accidental events, and
misreconstructed events, are difficult to remove completely via this
method.
In the reconstruction method in particle physics, additional physical
parameters that are not used to solve the kinematics are usually
needed to suppress such a background.
In MeV gamma-ray telescopes using the reconstruction of Compton
scattering, residual physical parameters are considered necessary
\citep{Tanimori2017}.
We have good examples.
COMPTEL successfully opened MeV gamma-ray astronomy by detecting
approximately 30 celestial objects using a CC \citep{Schonfelder1993,
 Schonfelder2000}, although the actual sensitivity was degraded by
25\% compared with that expected before launching
\citep{Schonfelder2004}.
According to the experience of SMILE-2+, this success is mainly
attributed not to the imaging of the CC method but to the efficient
background rejection based on the time of flight (ToF), the pulse
analysis of the forward liquid scintillator, the reduction of the FoV
(limited to 30$\degree$), and the adoption of a light material as a
scatterer of the forward detector (FD).
In principle, the CC cannot distinguish correctly reconstructed
Compton scattering events (signal) from misreconstructed background
events, owing to the incompleteness of the event reconstruction.
COMPTEL adopted several tools to reduce the background; in particular,
the ToF was effective for removing most of the background coming from
the downward direction, which accounted for approximately 90\% of the
total background \citep{Schonfelder2004}.
Additionally, the pulse-shape analysis provided a good reduction of
the neutron background, and the adoption of a light material was
effective for reducing the accidental background to reduce the single
hit counts in the FD.
The narrow FoV was effective for reducing the large background coming
from outside the FoV, although it also reduced the detection area.
COMPTEL appears to be able to reduce the background by more than one
order of magnitude and thus detect the celestial objects.
However, there remains a large background exceeding that of cosmic
diffuse MeV gamma rays by approximately one order of magnitude, as
shown in Fig.~\ref{fig2} of \citep{Kappadath}.
Thus, the study of COMPTEL indicates that additional parameters,
except for those needed to solve the kinematics of Compton scattering,
must be employed in MeV gamma-ray telescopes.

The measurement of the track of the recoil electron provides
additional parameters such as the energy loss rate ($d$E/$d$x) of the
particle track (Figs.~\ref{fig7} and \ref{fig19}), the scattering
angle $\alpha$ between the recoil electron and the scattered gamma
rays in Compton scattering (Figs.~\ref{fig1} and \ref{fig4}), and the
event topology, as shown in Fig.~\ref{fig8} \citep{Tanimori2017,
 Tanimori2015}.
The correlation between $d$E/$d$x and the energy of the tracks is
useful for particle identification, enabling to distinguish recoil
electrons from cosmic rays, high-energy electrons, and neutrons.
The angle $\alpha$~can be determined using the kinetic form of the
kinematic equation and from the measurements of the directions of the
recoil electrons and scattered gamma-rays.
Then, the two angles can be compared to determine whether the gamma
direction calculated via the kinematical equation is correct or false,
which is referred to as the kinematic test.
Furthermore, an event topology enables easy determination of whether
the measured event is due to Compton events caused by gamma rays by
requiring only one fully contained track in the tracking instrument,
as shown in Fig.~\ref{fig8}.

\begin{figure}[htbp]
 \begin{center}
\includegraphics[width=3.83333in]{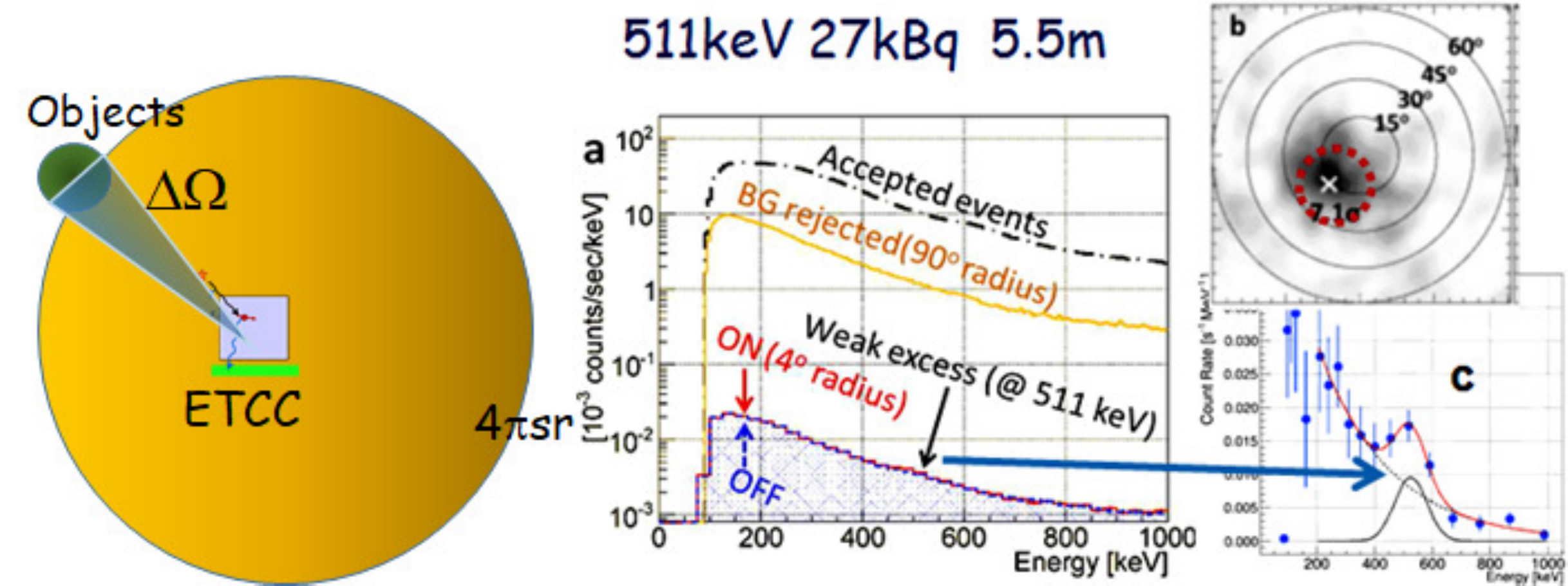}
 \end{center}
 \caption{Schematic explanation of background reduction by the
 PSF. (right) Energy spectra of a super-weak point source (27 kBq
 \textsuperscript{22}Na) at a distance of 5.5~m. (b) Image observed
 by the SMILE-II ETCC with 7.1$\sigma$. Adapted from \citep{Tanimori2015}. }
 \label{fig9}
\end{figure}

From the results of the ground experiments (Fig.~\ref{fig9}) and two
balloon experiments, $d$E/$d$x and the event topology were confirmed
to be more efficient than the ToF in COMPTEL without the loss of the
true signal events \citep{Takada2011}.
The combination of the two reduced the background by more than two
orders of magnitude at an altitude of $>$35 km to maintain a wide FoV
of $>$3 sr.

\begin{figure}[htbp]
 \begin{center}
\includegraphics[width=2.66769in]{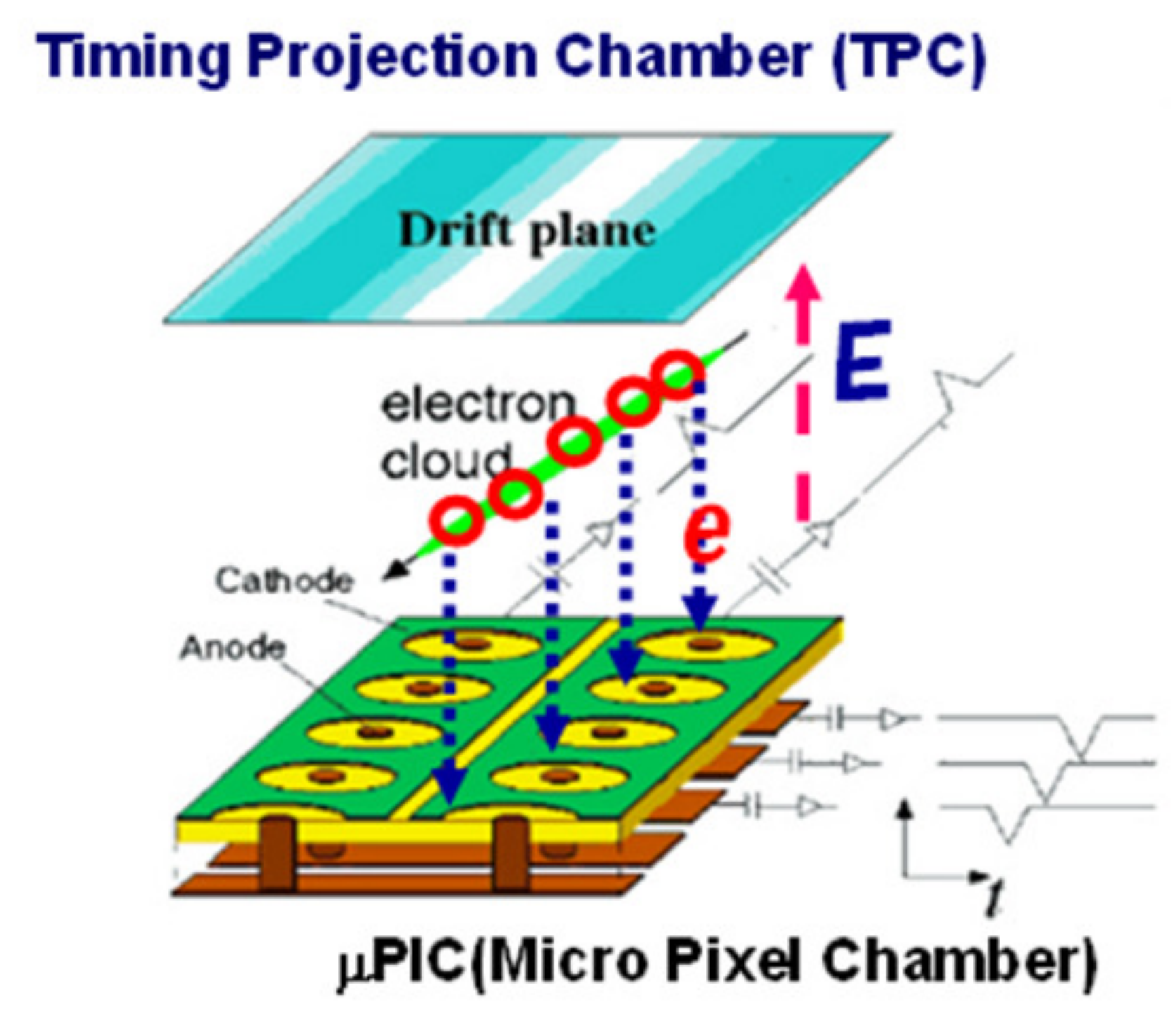}
 \end{center}
 \caption{Schematic of the $\mu$PIC and the operating principle of the TPC using $\mu$PIC.
 }
 \label{fig10}
\end{figure}

To realize the aforementioned requirements for almost all recoil
electrons ranging from a few kiloelectronvolts (keV) to several MeV,
an ideal three-dimensional (3D) tracking device such as a cloud
chamber that provides a fine image of beta-decay electrons is
necessary.
A gaseous TPC enables the measurement of the 3D image of such a fine
track of the above energy region electrically by using a micropattern
gas detector (MPGD), as shown in Fig.~\ref{fig10} \citep{Tanimori2004},
where the 3D position of the track can be measured with a sub-mm
pitch.
Herein, such a TPC is referred to as a $\mu$TPC.
Because a 10-keV electron track runs along 5 mm in 1 atm Ar gas, $>$10
points in one track can be measured, whose pitch is similar to the
diffusion of the drift electron passing through a distance of a few
tens of centimeters in the $\mu$TPC.

\subsection{Estimation of sensitivity of ETCC in MeV gamma astronomy}
\label{subsec:sensitivity:etcc}
 
For celestial objects, the sensitivity of telescopes is generally
calculated by knowing the effective area, proper PSF, and background
flux \citep{Tanimori2015}.
The background comprises the flux of cosmic diffuse gamma-rays (CDG),
which has
a celestial origin and is the minimum background that is never
removed; albedo gamma rays from the atmosphere; and instrumental
radiation generated by cosmic rays.
First, let us consider the minimum effective area and worst PSF to reach
a sensitivity of 1 mCrab during the observation time of 10$^{6}$ s for
the ideal case in which all background except CDG could be removed.
For a PSF of a few degrees, a relatively small effective area of a few
100 cm$^{2}$ is needed because the typical fluxes of celestial gamma
rays in the MeV region are three orders of magnitude stronger than
those in the GeV region.
If the PSF was \textless10$\degree$, an effective area 100 times
larger would be necessary.
Considering the limitations of the size and weight of middle-class
satellites, a possible scale of the effective area for the MeV gamma
telescopes is estimated to be a few to several hundred cm$^{2}$ for
1-MeV gamma-rays at maximum.
Thus, a proper PSF of a few degrees and complete background rejection
should be achieved simultaneously for all MeV gamma sky surveys,
aiming for a sensitivity of \textless1 mCrab \citep{Tanimori2015}.

\begin{figure}[htbp]
 \begin{center}
\includegraphics[width=4.in]{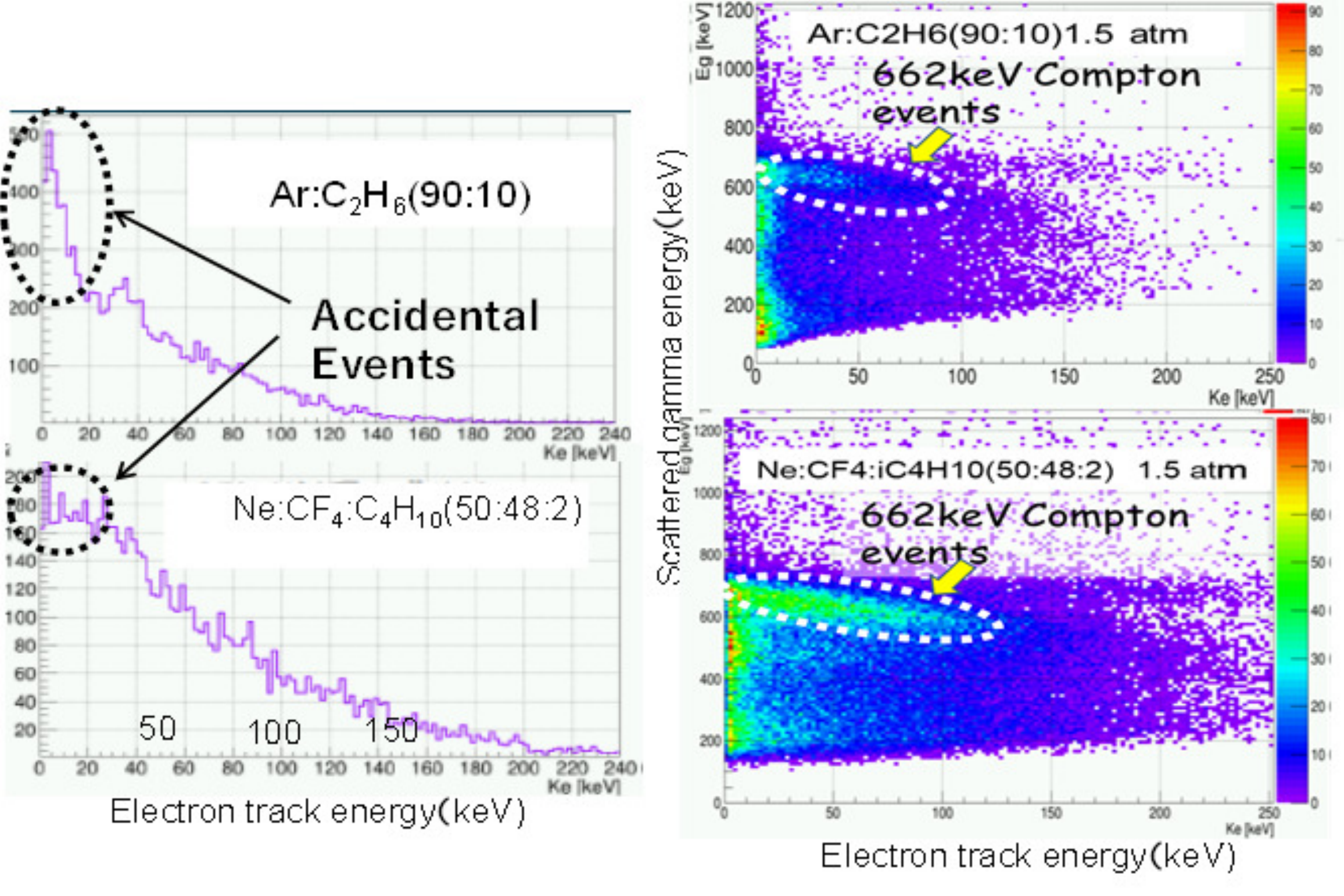}
 \end{center}
 \caption{(left) Accidental events on the energy spectra of the
 recoil electron in the TPC for Ar and Ne-CF\textsubscript{4}, as
 well as the Compton events in the relationship of the scattered
 gamma energies and the recoil electrons for both gases. }
 \label{fig11m}
\end{figure}

Almost all proposals for MeV gamma astronomy using CCs mention the
effective area and the ARM
 (Angular Resolution Measure)
as an angular resolution or the effective
PSF; however, they hardly describe the method of estimating the sensitivity
from the effective area and ARM quantitatively.
As mentioned previously, the image of the CC is a nonlinear system.
In a linear system, the event density of the imaging plane increases
linearly as the number of photons entering the FoV increases, whereas
the event density increases almost with the square of the number of
photons because each event annulus surely crosses several other events
whose number increases linearly as the events increase, as shown in
Fig.~\ref{fig5}.
The following general formula 
for estimating the sensitivity in astronomy must not be used simply for
the CC.
In astronomy, the significance of a source is 

\begin{equation}
 \propto \frac{A_{\eff}\cdot S}{\sqrt{A_{\eff}\cdot\left(S+B\cdot\theta^2\right)}},
\end{equation}
which can be approximated in the background-dominated case to
\begin{equation}
 \propto \frac{A_{\eff}\cdot S}{\theta\sqrt{A_{\eff}\cdot B}},
\end{equation}
where $S$ is the signal intensity, $B$ the background flux, $A_{\eff}$
is the effective area, and $\theta$ is the half-power radius of the
PSF.

\begin{figure}[htbp]
 \begin{center}
\includegraphics[width=2.5in]{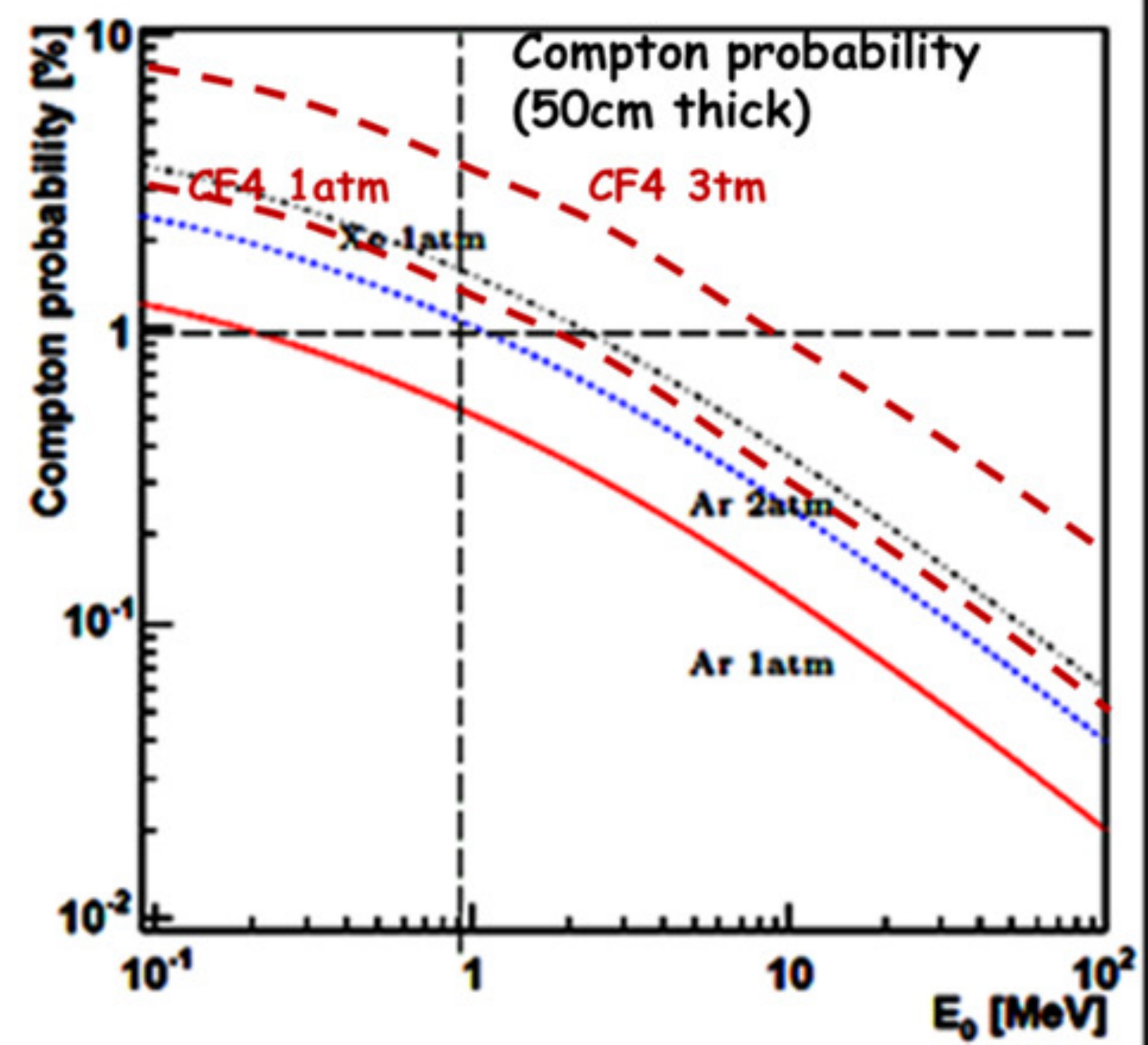}
 \end{center}
 \caption{Energy dependences of the scattering probabilities of Ar, Xe, and CF\textsubscript{4} gases with normal and 3-atm pressures.
 }
 \label{fig12m}
\end{figure}

A gaseous tracking device for measuring the recoil electron direction
is a unique solution.
How can a gaseous detector provide the required effective area in a
1-m$^{3}$ cubic volume? The key point of Compton scattering is that it
is proportional to the atomic number or electron density, in contrast
to the photoelectric effect.
Molecular carbonic gas is a good choice for the scatterer.
CF\textsubscript{4} has 42 electrons in one molecule, which is nearly
three times that of Si, and its low atomic number significantly
suppresses the photo absorption of gamma rays in the scatterer, which
is the main reason for the accidental background \citep{Tanimori2017,
 Tanimori2015}.
Fig.~\ref{fig11m} shows the good enhancement of the Compton events and
suppression of the accidental events due to the use of
CF\textsubscript{4} gas measured by a (10 cm)$^{3}$ cubic ETCC in 2015.

\begin{figure}[htbp]
 \begin{center}
\includegraphics[width=2.75in]{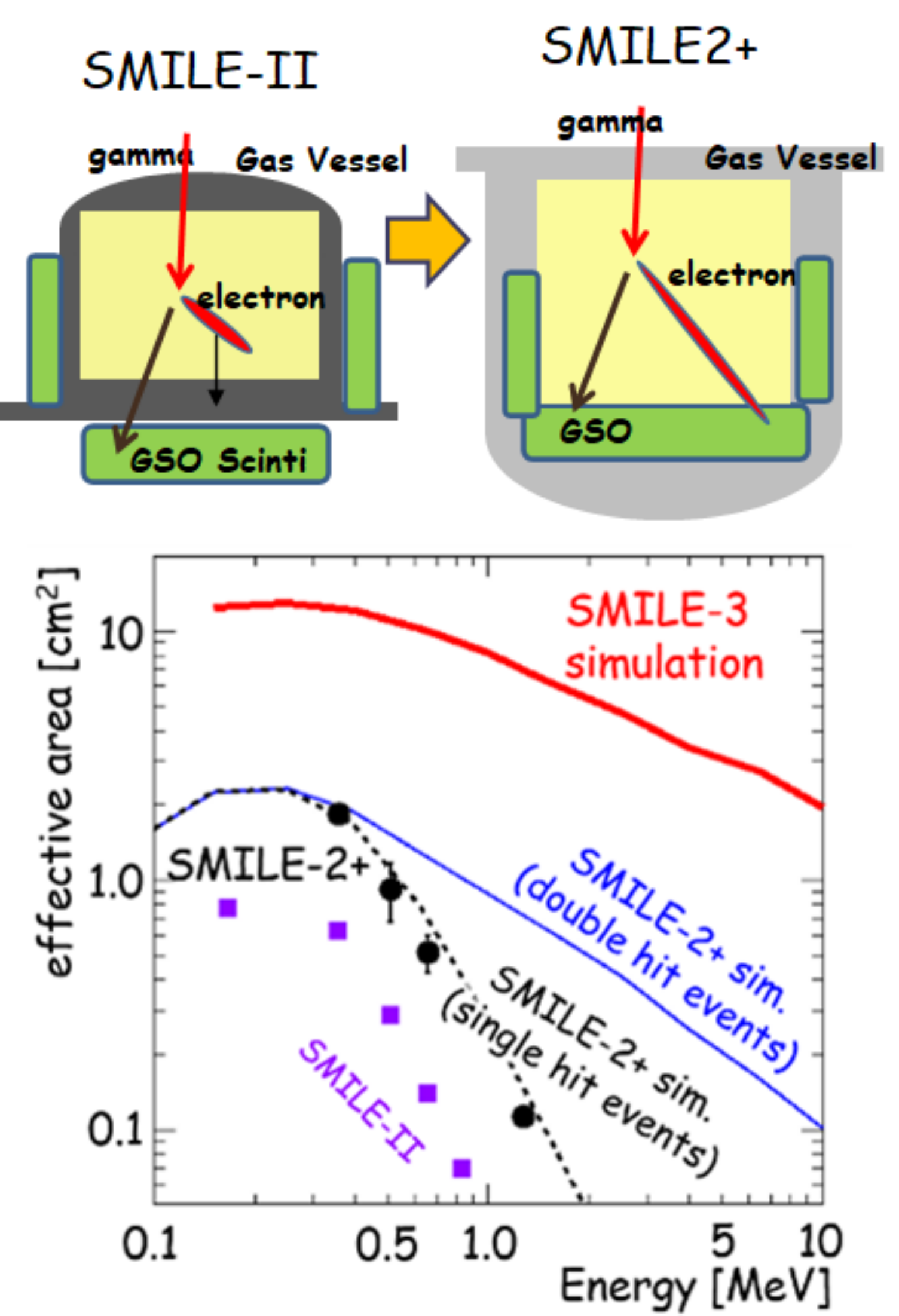}
 \end{center}
 \caption{(top) Explanation of how to detect a high-energy recoil
 electron via SMILE-2+. (bottom) Energy dependences of the effective
 areas for SMILE-II, SMILE-2+, and SMILE3. }
 \label{fig13}
\end{figure}

\begin{figure}[htbp]
 \begin{center}
\includegraphics[width=2.5in]{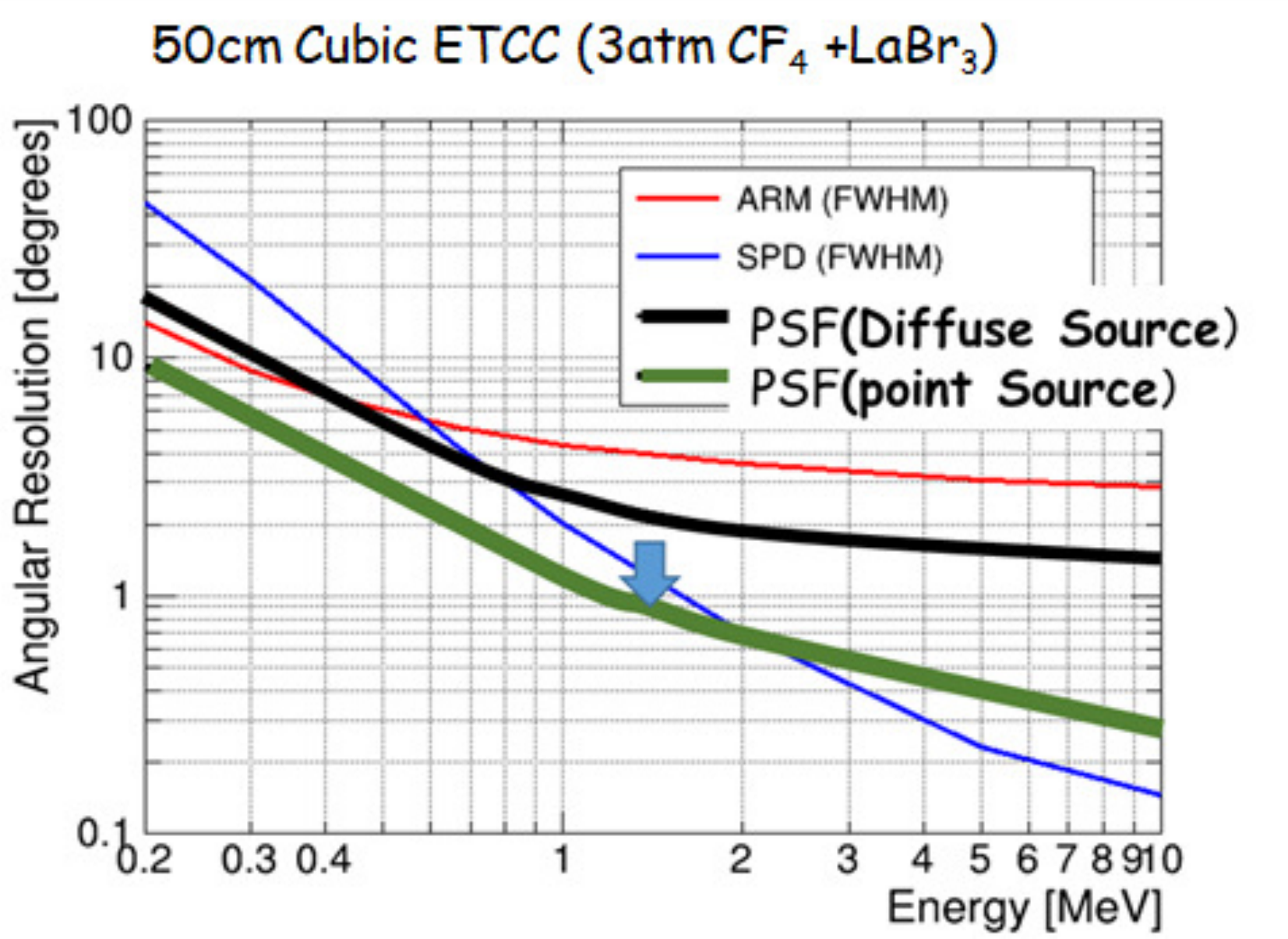}
 \end{center}
 \caption{Energy dependence of angular resolution-related
 parameters. Red and blue lines indicate the ARM and SPD,
 respectively. The PSF is indicated by a black thick line; assuming
 a point source, the PSF is improved by weighting the SPD, as
 indicated by the thick green line.
 Adapted from \citep{Mizumura}. }
 \label{fig14}
\end{figure}

COMPTEL uses a liquid scintillator as the scatterer, which comprises
mainly carbon and hydrogen \citep{Schonfelder2004}.
A (50 cm)$^{3}$ cubic CF\textsubscript{4} gas with 3 atm gives an
effective area of 110 cm$^{2}$ for 1-MeV gamma-rays, as estimated from
Fig.~\ref{fig12m}, and a tracking device with a 50-cm$^{2}$ effective
area is easily realized using TPC technology.
The TPC can measure the 3D position of a charged particle inside a
large gas volume using only one 2D MPGD.
A (50 cm)$^{3}$ cubic TPC is relatively small and obtains a position
resolution of a few hundred micrometers without the magnetic field to
suppress the diffusion of the drift electron in the gas volume.
If 40\% of the Compton scattering is detected in the TPC, 4$\times$
(50 cm)$^{3}$ cubic TPCs provide an effective area of $\sim$200 cm$^{2}$
at 1 MeV.
Sub-mm sampling with a size of 50 $\times$ 50 cm$^2$ is common for MPGD,
and the gas amplification of 10$^{4}$ in the TPC enables the use
of simple, light, and low-power readout electronics.
Thus, the gaseous TPC reduces the amount of readout electronics by
more than one order of magnitude compared with the multi-layer solid
detector.
Furthermore, there is no material except the gas in the TPC gas
volume; thus, scatters that took place in the scatterer can be
selected by measuring the electron track inside the gas volume.
A multilayer structure requires many materials to be
supported.
cooling, and electrically handling the scatterer devices
inside and near the FD of the CC, which causes a huge background that is
not distinguished from real events scattered in the scatterer.

\begin{figure}[htbp]
 \begin{center}
\includegraphics[width=2.75in]{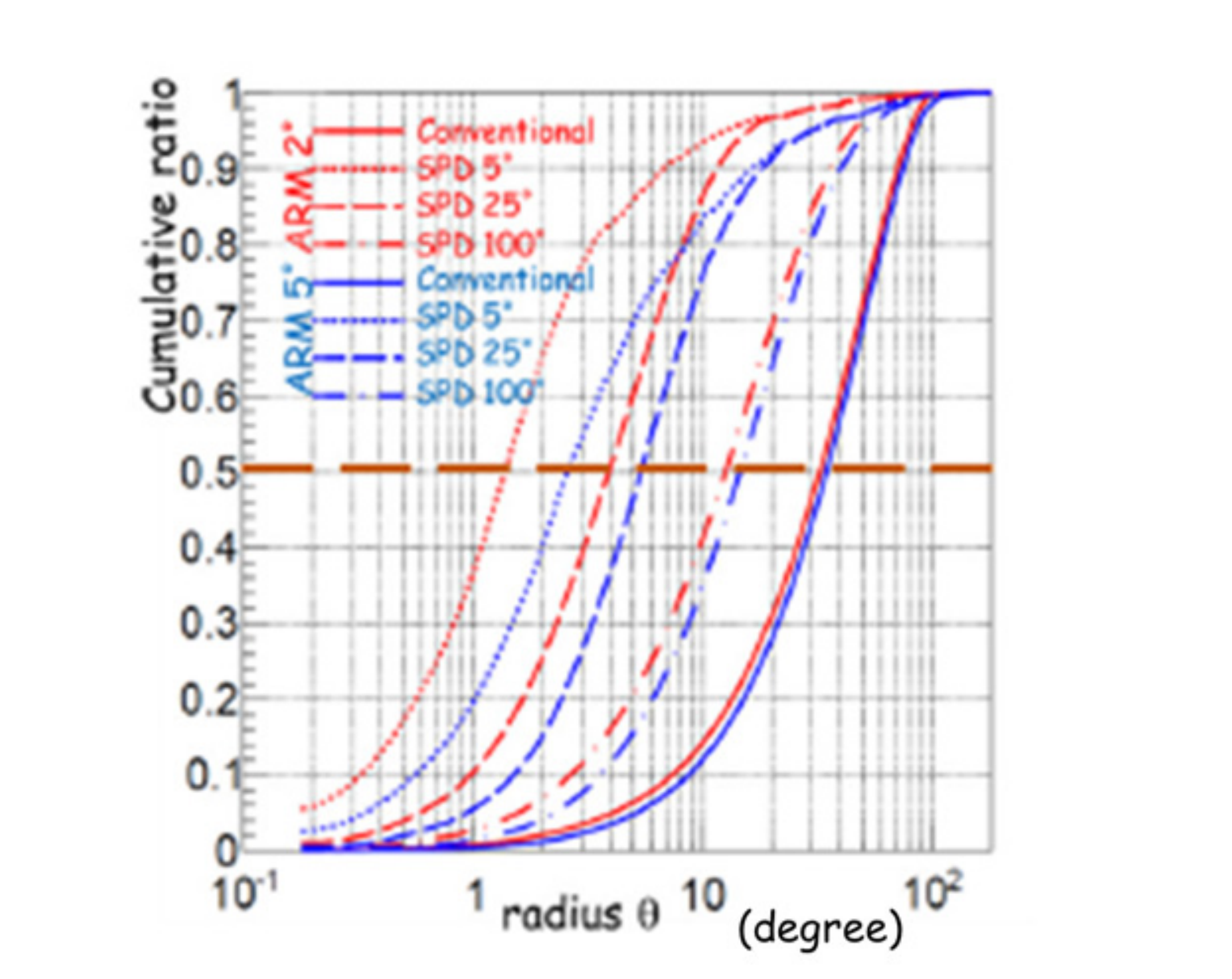}
 \end{center}
 \caption{Cumulative ratios of events within a radius of $\theta$
  (50\% corresponds to the PSF). 
Adapted from \citep{Tanimori2015}. }
 \label{fig15}
\end{figure}

Additionally, the TPC provides a moderate energy resolution for the
electron track at room temperature: 20\%, 10\%, and a few \% for 10,
50, and $>$100 keV, respectively.
By combining a good scintillator or a heavy semiconductor as the
absorber surrounding the TPC, a good energy resolution of 2\%--5\% at
662 keV was obtained at room temperature.

For the use of gas as a scatterer, there is one serious problem: a
recoil electron with an energy higher than several hundreds of keV
penetrates the TPC and significantly reduces the effective area for
gamma-rays above a few MeV.
To overcome this problem, all absorbers were moved inside the TPC in the
SMILE-2+ experiment \citep{Mizumura,Takada2021}, as shown in
Fig.~\ref{fig13}.
Recoil electrons with energies of $>$1 MeV provide a good sub-degree
SPD (Scatter Plane Deviation)
for gamma rays above a few MeV, as shown in Fig.~\ref{fig14},
while the ARM is saturated to a few degrees above a few MeV
\citep{Mizumura}.

\subsection{How to obtain a good PSF}
\label{subsec:psf:etcc}

\begin{figure}[htbp]
 \begin{center}
\includegraphics[width=2.52917in]{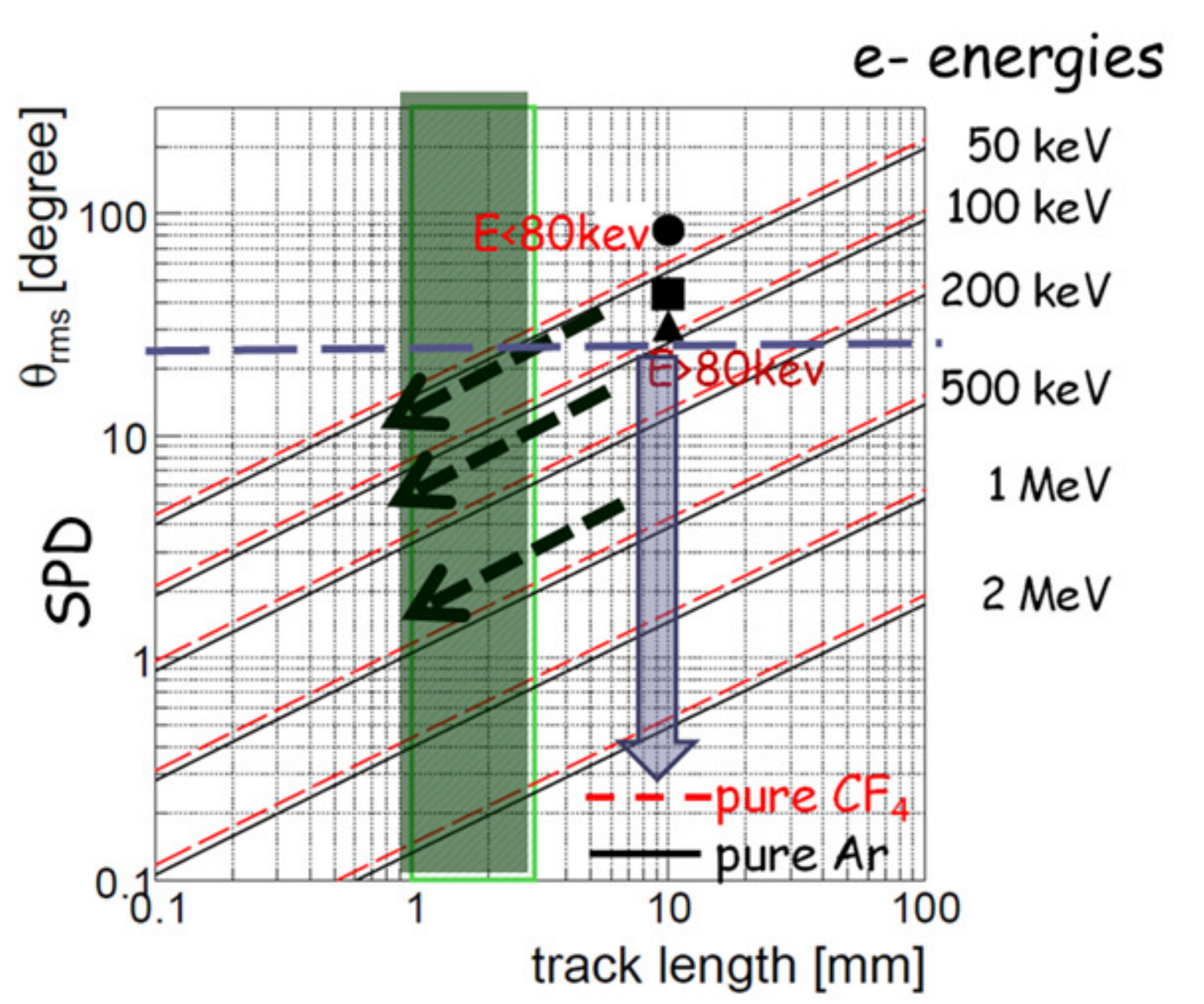}
 \end{center}
 \caption{Relationship of the multiple scattering (SPD) of a recoil
 electron and the length of tracking in 3-atm CF\textsubscript{4}
 and Ar gases for several energies of recoil electrons.
Adapted from \citep{Tanimori2015}. }
 \label{fig16}
\end{figure}

The PSF is the most significant factor for attaining a good
sensitivity.
In Compton scattering, the PSF is determined equivalently by the two
angular resolutions of the elevation and azimuth in general, as shown
in Figs.~\ref{fig14} and \ref{fig15}.
The former is the ARM, and the other is a function of the SPD
\citep{Tanimori2017, Tanimori2015}.
To obtain a good PSF, the two angular resolutions must be similar.
The ARM depends on the energy resolution of the telescope, and as the
Compton scattering angle increases, the ARM becomes worse.
Although the theoretical ARM of the forward scattering region is
better (by a few degrees) than those in larger scattering regions, the
ARM is degraded by geometrical errors owing to the position
resolutions of the scatterer and absorbers and systematic errors of
the instrument structure.
Because the position resolutions of semiconductors or gaseous TPCs are
sub-mm, the real ARM depends on the distance between the hit points of
the scatterers and absorbers, which is the instrument size.

For COSI with Ge \citep{Kierans:2019aqz} and SDGs
 (Soft Gamma Ray Detector) on Hitomi satellite
with cooling Si and CdTe \citep{Watanabe:2014ojx}, an ARM of
6$\degree$ at 511 and 662 keV were reported, respectively, with energy
resolutions of \textless1\% and $\sim$2 \% at 662 keV, but for the
(30 cm)$^{3}$ cubic ETCC of SMILE-II, an ARM of 5$\degree$--6$\degree$ at
662 keV was obtained using a GSO scintillator with an energy
resolution of 11\%.
A TPC with a Gd$_3$(Ga,Al)$_5$O$_{12}$(Ce) (GAGG) scintillator having
a good energy resolution of 5\% at 662 keV provides a good ARM of
4$\degree$ at 662 keV \citep{Tanimori2017, Tanimori2015}.
However, for a good PSF, the improvement of the SPD is far more
important considering the present resolution.
Because the SPD is the angular uncertainty of the Compton scattering
annulus, it improves the resolution of the PSF by roughly a factor of
the ratio of the scattering angle to 180$\degree$
\citep{Tanimori2017,Mizumura}.
For example, an SDP of 25$\degree$ (full width at half maximum, FWHM)
with an ARM of 5$\degree$ (FWHM) gives a 3$\degree$ HPR (Half power
radius) as the PSF,
which is similar to the PSF of EGRET at 100\,MeV \citep{NASA:EGRET},
as shown in Fig.~\ref{fig15}.
The SPD is mainly determined by the multiple scattering of recoil
electrons in the scatterer.
As shown in Fig.~\ref{fig16}, a good PSF requires an SPD smaller than
20$\degree$, which is obtained by a few 100 $\mu\textrm{m}$ 3D
sampling of a gaseous TPC.

This requirement was already attained by the MPGD TPC used in the
SMILE project.
For Si, 100-nm 3D sampling is required, which is difficult, even in
the near future.
Thus, only gaseous~$\mu$TPC has the ability to provide both a
sufficient PSF and an effective area to attain sub-mCrab sensitivity
in a satellite-scale detector, which is an essential reason for using
a gas detector for MeV gamma-ray imaging spectroscopy
\citep{Tanimori2017, Mizumura}.

\subsection{Development of ETCC}
\label{subsec:development:etcc}

\begin{figure}[htbp]
 \begin{center}
\includegraphics[width=4.in]{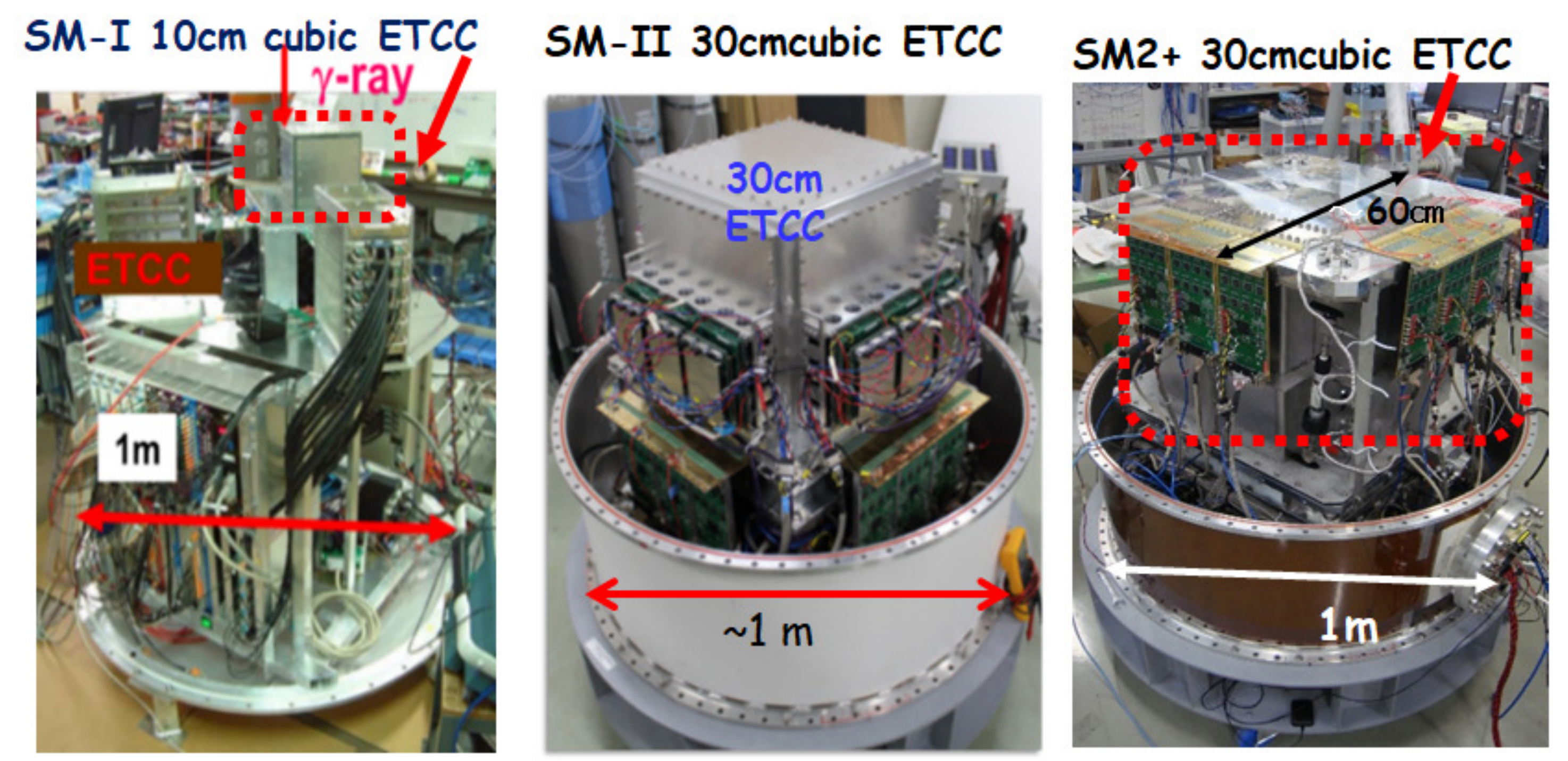}
 \end{center}
 \caption{Photographs of SMILE-I, SMILE-II, and SMILE-2+.
 }
 \label{fig17}
\end{figure}

The ETCC is a unique instrument that satisfies all the aforementioned
requirements for a breakthrough in MeV gamma-ray astronomy.
As shown in Fig.~\ref{fig1}, the ETCC consists of an FD as a scatterer
of gamma-rays and a backward detector (BD), which functions as a
calorimeter for the scattered gamma-rays.
A gaseous TPC is used as the FD, which is based on MPGDs, to measure
the 3D tracks of recoil electrons, and pixel scintillator arrays
(PSAs) with heavy crystals (at present,
Gd\textsubscript{2}SiO\textsubscript{5}:Ce, GSO) are used for the BD.
Ar-based gas was used for TPCs with an energy resolution of
$\sim$30\% at 5.9 keV because the gas amplification of
CF\textsubscript{4} gas requires a higher voltage than that of Ar gas
\citep{Tanimori2017}.
Although a micropixel gas counter ($\mu$PIC) \citep{Ochi:2000js} was
originally developed to provide a higher gas gain of
$>10^{4}$ in MPGDs, which are mostly operated
with a gain of \textless10$^{4}$, the $\mu$PIC was operated at a
lower gain of $6 \times 10^{3}$ to avoid discharge damage.
Additional gain was provided by a GEM with a gain of several units,
and a total gain of $1.5 \times 10^{4}$--$2 \times 10^{4}$ obtained
routinely.
This two-step gain system has provided stable operation for more than
10 years. Recently, a new $\mu$PIC was developed using a glass
substrate instead of a polyimide one \citep{Abe}
and this new $\mu$PIC is not
destroyed by discharges and provides a higher gain than the present
$\mu$PIC by a factor of $>$2.
Operations with CF\textsubscript{4} is planned, so as to attain an
effective area of 10 cm$^{2}$ for a (30 cm)$^{3}$ cubic TPC.

The energy resolution of the ETCC with GSO scintillators using a
photomultiplier tube was 11\% (FWHM) at 662 keV.
New PSAs with GSO + SiPM and GAGG + SiPM can yield 8\% and 5\% (FWHM)
at 662 keV, respectively.

\begin{figure}[htbp]
 \begin{center}
\includegraphics[width=4.5in]{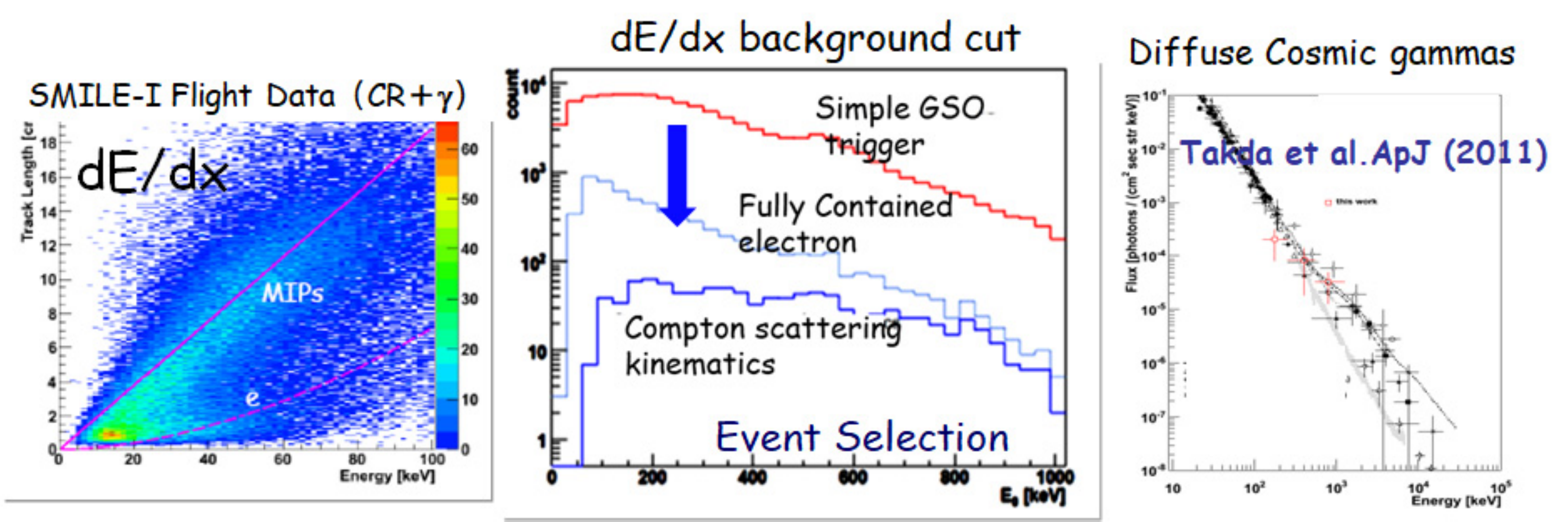}
 \end{center}
 \caption{(left) $d$E/$d$x scatter plot of SMILE-I balloon
 observation; (center) variations of spectra due to event electrons
 in SMILE-I observation; (right) spectra of diffuse cosmic gamma-rays
 obtained via SMILE-I. }
 \label{fig18}
\end{figure}

In 2004, for the first time the successful full electron
tracking in a laboratory experiment with a small
(10 cm)$^{3}$ cubic ETCC was reported \citep{Tanimori2004, Takada2005}.
Then the ``Sub-MeV gamma ray Imaging Loaded-on-balloon Experiment''
(SMILE) project was conducted with the improved (10 cm)$^{3}$ cubic ETCC
(SMILE-I in Fig.~\ref{fig17}) to measure the diffuse cosmic MeV
gamma-rays via a balloon-borne experiment in 2006 \citep{Takada2011}.
An excellent particle-identification ability according to the
$d$E/$d$x of an electron track in the gas was demonstrated as shown in
Fig.~\ref{fig18}, from which the background level was 
reduced by a factor of $\sim$3. 

A (30 cm)$^{3}$ cubic ETCC (Figs.~\ref{fig1} and
\ref{fig17})
was then developed
to achieve an effective area of 1 cm$^{2}$ at 300 keV
(Fig.~\ref{fig13}) for the detection of celestial MeV-gamma-ray
objects such as the Crab with a balloon experiment in the Northern
Hemisphere (SMILE-II) \citep{Tanimori2015}.
In this ETCC, the tracking efficiency in the TPC is improved
significantly (from 10\% to 100\%) compared to SMILE-I owing to the improved
readout electronics and algorithm.
This provides better noise reduction of $d$E/$d$x and good angular
resolutions of 5.9$\degree$ (FWHM) at 662 keV, which is consistent
with that calculated using the detector energy resolution
\citep{Mizumoto}.

\begin{figure}[htbp]
 \begin{center}
\includegraphics[width=4.in]{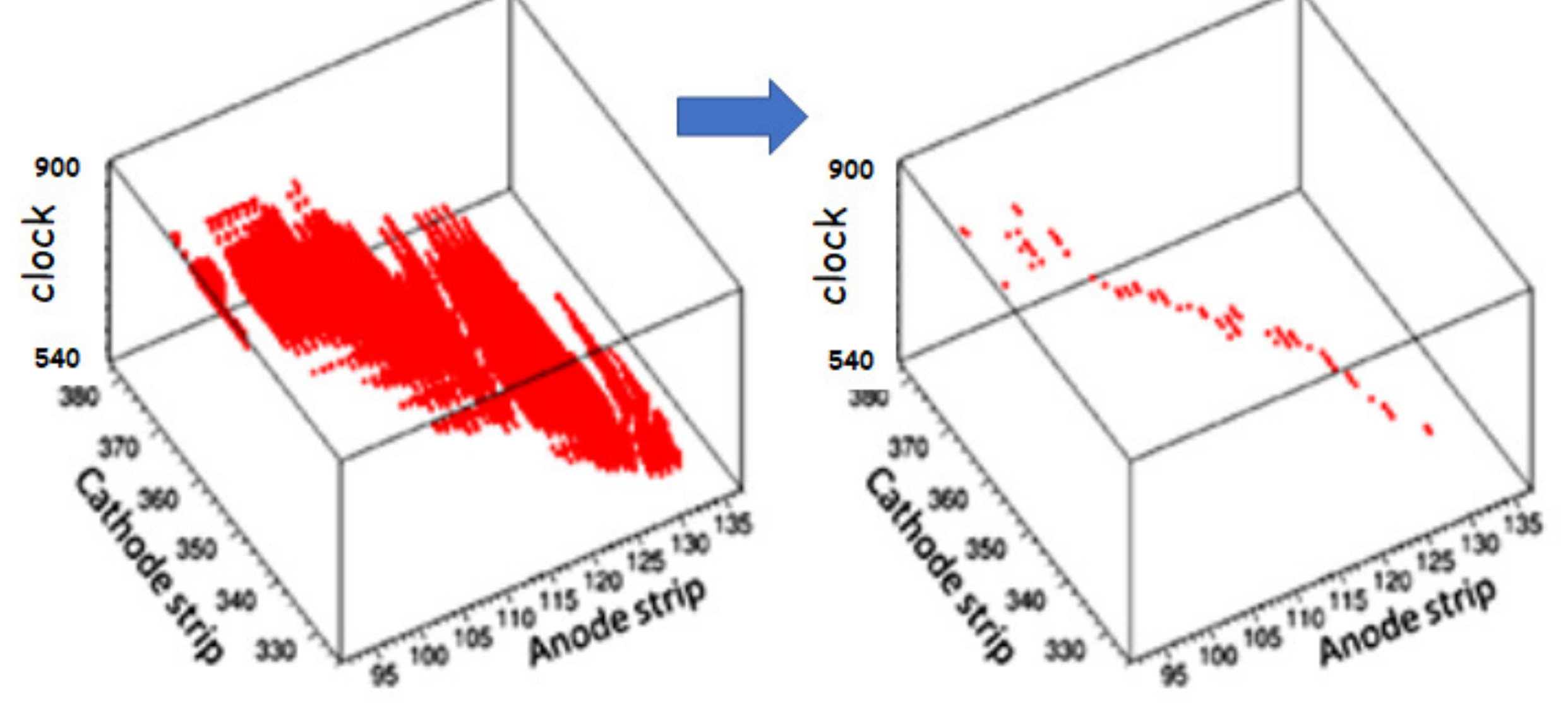}
 \end{center}
 \caption{3D electron-tracking images with the ambiguity from the
 multi-hits 
 (left) before and
 (right) after the correction of time walk.
 Adapted from \citep{Tanimori2015}. }
 \label{fig19}
\end{figure}

For the ETCC, obtaining the direction of the recoil electron is crucial.
In the $\mu$PIC of the TPC, orthogonal strips of the X and Y coordinates
were used to significantly reduce the number of readout channels from
the pixel readout.
However, this readout method causes a well-known left--right
uncertainty in tracking multiple hits simultaneously as shown in
Fig.~\ref{fig19} \citep{Tanimori2015}, which degrades the SPD
considerably to 200$\degree$ (FWHM) with SMILE-I.
In the new readout electronics of SMILE-II, the pulse width of each
electrode in the $\mu$PIC is recorded as the timing width over the
threshold (TOT) \citep{Mizumoto}, which provides a coarse charge
deposit at each hit point.
In 2015, using the TOT, the coincidence width between the X and Y
strips could be reduced from 10 to $\sim$1 ns and the SPD improved to
$\sim$100$\degree$ (FWHM), thus reducing the uncertainty in tracking
\citep{Tanimori2015}.

Consequently, a half-power radius of 25$\degree$ for the PSF at 662
keV was obtained
\citep{Tanimori2015}.

\subsection{SMILE-2+ balloon experiment}
\label{subsec:smile2}

SMILE-2+ was improved from SMILE-II for a sufficient sensitivity
estimated according to the PSF for the Galactic Center (GC) diffuse
gamma-ray flux \citep{Tanimori2020, Takada2021}.
SMILE-II was the first (30 cm)$^{3}$ cubic ETCC with 1-atm Ar gas, and
PSAs were set outside the gas vessel of the TPC.
SMILE-II was designed to use the ARM as a PSF for detecting Crab with
$>$3$\sigma$ via several hours of observation in the Northern
Hemisphere \citep{Tanimori2015}.
However, given the importance of the proper PSF for the estimation of
sensitivity \citep{Tanimori2017}, SMILE-II was redesigned to SMILE-2+
to realize the same sensitivity as that expected for SMILE-II, taking
into account that the flux of Crab is reduced by half owing to the
smaller zenith angle of 45$\degree$ in the southern-sky observation
\citep{Takada2021}.
The effective area of SMILE-2+ was increased three times at 511 keV and
several times above 1 MeV with Ar gas at 2 atm and the use of a GSO of
double thickness in the bottom pixel scintillation arrays (PSAs).
Additionally, because a fully contained electron in the TPC restricts
the energy range of gamma-rays to \textless1 MeV, all the PSAs were set
inside the TPC to detect the higher-energy recoil electron, as shown
in Fig.~\ref{fig13}.

A one-day observation was conducted with a large FoV ($>$3 sr) for the
entire southern sky during the JAXA balloon launching at Alice Springs
of Australia on April 7--8, 2018 \citep{Takada2021}.
The balloon was successfully flown at an altitude of 37--39 km for
30~h, and the southern sky was observed, including the GC, half of the
Galactic disk, Crab, Cen-A, and the Sun.

\subsection{Analysis for background reduction}
\label{subsec:analysis:background:reduction}

$4 \times 10^{7}$ triggered events were recorded and
finally reduced to 10$^{5}$ events (by more than two orders of
magnitude) with the requests of the fully contained and good $\dd E/\dd x$
for the recoil electron in the TPC.
Here, the clear event topology enabled the efficient identification of
Compton scattering.
The number of final events is consistent with that estimated from the
SMILE-I results, indicating a signal-to-background ratio of $>$1.
This is clearly confirmed by the obvious enhancements of the detected
gamma flux at $\sim$10\% during the passage of the GC through the FoV,
for both low- and high-energy events, as reported in
\citep{Takada2021}.
The ratio is consistent with previously reported data, and as a
result, the final data contained only a few 10\% of the instrumental
background.
Until now, for all observations of MeV gamma-rays done before SMILE-2+, the
background was approximately 1--2 orders of magnitude higher than the
signal.
For SMILE-2+, because 3/5 of the whole sky was observed via bijection
imaging, the OFF-region could be set simply with both the GC and the
disk outside the FoV, and the significance showed enhancements of the
GC and disk were observed.

\begin{figure}[htbp]
 \begin{center}
\includegraphics[width=3.in]{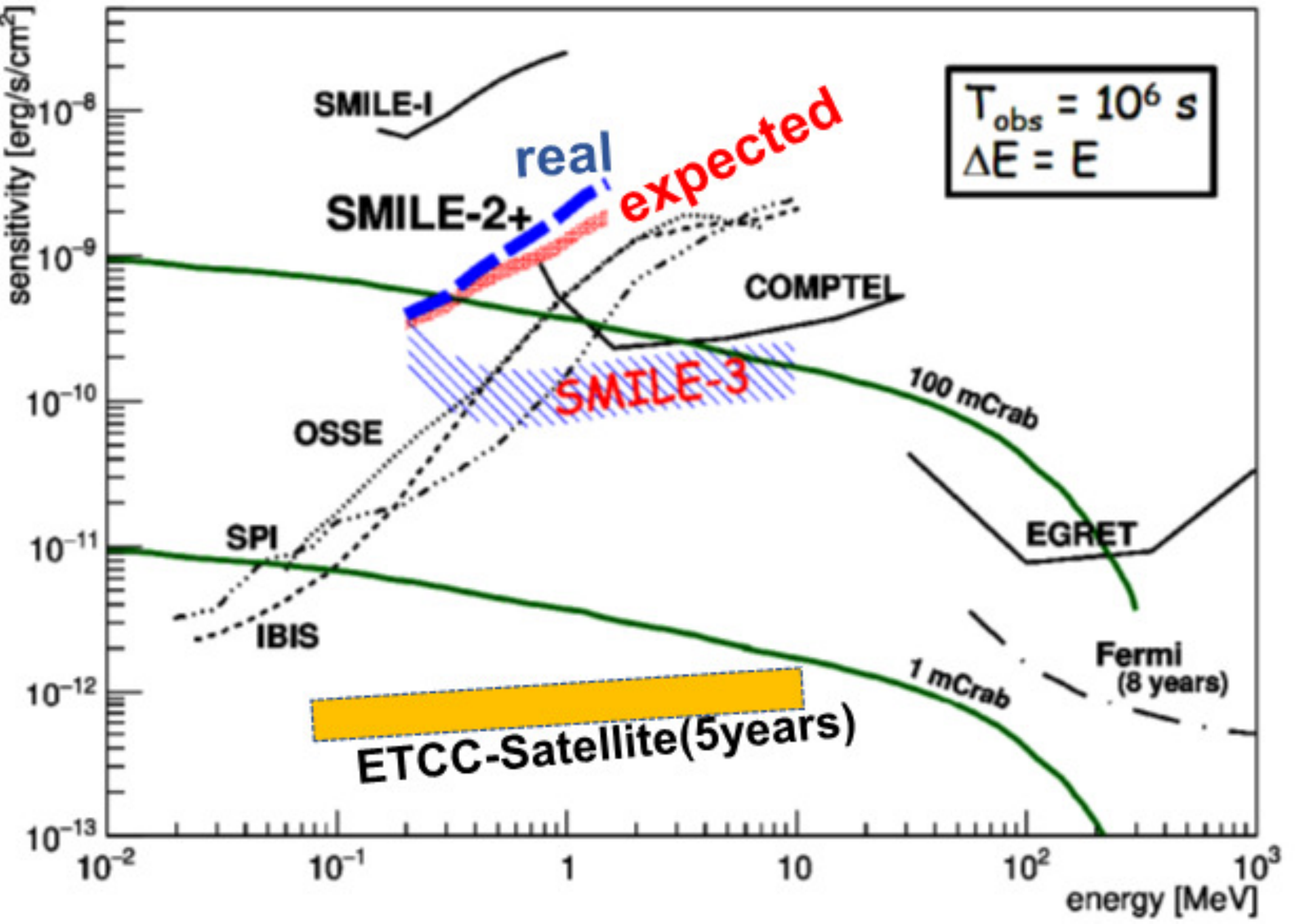}
 \end{center}
 \caption{Observed and expected sensitivities of SMILE-2+ and expected sensitivities of SMILE3 and ETCC-satellite.
  Adapted from \citep{Hamaguchi, Takada2021}.
 }
 \label{fig20}
\end{figure}

ETCC has been demonstrated to enable MeV gamma-ray astronomy in the
same manner as for astronomy at other wavelengths, i.e., bijection
imaging and complete background rejection.

\subsection{Future prospects}
\label{sec:etcc:future:prospects}

The next step will be the construction of SMILE3, a 40~cm $\phi\times$
30~cm cylindrical ETCC with 3-atm CF\textsubscript{4} gas and a
3-coordinate~$\mu$PIC, which give a 10-cm$^{2}$ effective area and a
good PSF of $\sim$5$\degree$ \citep{Takada2020, Takada2021}.
Thanks to the excellent rejection of background events, this
detector
will enable the study of the MeV gamma sky with a sensitivity several
times higher than that of COMPTEL for only one-month-long balloon
observation, as shown in Fig.~\ref{fig20}.

Such a detector on a space mission would yield a sensitivity enabling
the detection of faint sources at the sub-mCrab level, as shown in
Fig.~\ref{fig20} \citep{Tanimori2015}.
A few hundred cm$^{2}$ and a PSF of a few degrees are required.
Such a fine PSF requires a good SPD of $\sim$10$\degree$, which is
only possible when gas is used as a scatterer.
With molecular gases, the needed effective area can be obtained from a
1-m$^{3}$ sensitive volume.
A proposal of such an ETCC on a satellite mission was submitted to the
2020 NASA Decadal Survey \citep{Hamaguchi}.

\section{TPCs as pair telescopes}
\label{sec:pair}

After the first exploration of the gamma-ray sky by COS-B and
EGRET, it was recognized that further improvement in the crowded
and/or bright part of the sky, such as the galactic plane, and further
improvement at low energies would imply the removal of the high-$Z$
converters and the use of low-multiple-scattering active targets, such
as those consisting of gas drift chambers \citep{Mukherjee:1994zr}.
Later, and for similar reasons, the potential of such telescopes for
the gamma-ray polarimetry, the measurement of the fraction and
direction of the linear polarization of the incoming radiation, was underlined 
\citep{Bloser:2003gb}.

\subsection{Polarimetry with pair conversions and multiple scattering}
\label{subsec:pair:polarimetry}

Due to the $J^{PC} = 1^{--}$ nature of the photon, the reduced
differential cross section of the interaction of photons with a
charged particle, as a function of the azimuthal angle, $\varphi$, 
that measures
the orientation of the final state in the plane orthogonal to the
photon direction of flight, has the form:

\begin{equation}
\frac{\dd \Gamma}{\dd \varphi} \propto 
(1 + A P \cos{[2(\varphi-\varphi_0)]}),
\label{eq:modulation}
\end{equation}

\begin{itemize}
\item
 $P$ is the fraction of the linear polarization of the photon beam;
\item
 $A$ is the polarization asymmetry of the conversion process;
 its value depends on the conversion process (pair, Compton,
 photo-electric) and on the photon energy.
\end{itemize}

A number of experimental effects affect the measurement of the
azimuthal angle, among which multiple scattering is certainly the
fiercest and has been one of the major obstacles to polarimetry with
the past and present pair-conversion gamma-ray telescopes.
The presence of a non-zero resolution $\sigma_\varphi$ on the
azimuthal angle changes eq. (\ref{eq:modulation}) to
\begin{equation}
\gfrac{\dd N}{\dd \varphi} \propto 
\left(
1 + A \, e^{-2 \sigma_\varphi^2} P \cos[2(\varphi - \varphi_0)]
\right),
\label{eq:modulation:MS}
\end{equation}
so the effective polarization asymmetry is afflicted with a dilution factor 
$D = A_{\text{eff}} / A = e^{-2 \sigma_\varphi^2} $. 
The classical calculation \citep{Kelner,Kotov,Mattox} of
$\sigma_\varphi$ is performed assuming the small-polar-angle
approximation,
the most probable value $\hat\theta_{+-} = E_0/E$ of the pair opening angle
$\theta_{+-}$, with $E_0 = 1.6\,\mega\electronvolt$ \citep{Olsen1963},
and 
an approximate expression for the R.M.S. multiple
scattering deflection
(eq. (34.16) of \citep{Zyla:2020zbs}),
$\theta_0 \approx p_0/p \sqrt{x/X_0}$
($p_0 = 13.6\,\mega\electronvolt/c$,
$p$ the track momentum, 
$x$ and $X_0$ are the material thickness traversed by the electron and
the radiation length).
Assuming an equal energy share, $p \approx E / (2c)$, we obtain
(eq. (15) of \citep{Bernard:2013jea})
$\sigma_\varphi = \sigma_0 \sqrt{x/X_0}$ with
$ \sigma_0 \approx 24\,\radian$.
A dilution of $D = 1/2$ would be reached after a propagation of
$\approx 110\,\micro\meter$ in silicon, for example, that is before
the leptons could even exit the conversion wafer in a silicon-strip detector
(SSD) active target.

The catastrophic exponential dependence of the dilution as a function
of thickness so obtained,
$D \approx \exp(- \sigma_0^2 x / (2 X_0))$
is not confirmed by the results
of full simulations, actually,
(Fig. 7 of \citep{Bernard:2019znc}, 
Fig. 3 of \citep{Eingorn:2015oga}).
At high thicknesses, the dilution is found to be much larger
(i.e. less degraded) for the full simulation, something which is the
consequence of the $\theta_{+-}$ distribution having a huge tail at
large values \citep{Olsen1963}.

An optimal fit (such as a Kalman filter \citep{Fruhwirth:1987fm}) can
be performed that takes into account multiple scattering and the space
resolution of each of the detector layers $\sigma$.
For homogeneous detectors, the single-track angular resolution at the
vertex is found to be
 $ \sigma_{\track} \approx (p/p_1)^{-3/4}$, with
$p_1 = p_0 \left({4 \sigma^2 l}/{X_0^3} \right)^{1/6}$,
$l$ being the longitudinal sampling (along the track).
Given the $1/E$ scaling of the $\theta_{+-}$ distribution
\citep{Olsen1963}, the induced dilution is then found to be higher
(less degraded) at low energies (Fig. 20 of \citep{Bernard:2013jea}).

\subsection{Past experimental achievements and future prospects}
\label{subsec:pair:achievements:prospects}

Gamma-ray beams are currently produced in the laboratory using the
Compton scattering of a laser beam on a high-energy mono-energetic
electron beam.
Gamma-ray pseudo-monochromaticity is achieved by selecting the
``Compton-edge'' with a cylindrical collimator on the forward axis.
The linear polarization of the incident laser beam is then transferred
almost exactly to the gamma-ray beam \citep{Sun:2011es}.

\subsection{HARPO}
\label{sec:harpo}

The HARPO project (Hermetic ARgon POlarimeter) has developed a gas TPC
prototype \citep{Bernard:2018jql} and characterized it on the
gamma-ray beam of the BL01 beam line at LASTI \citep{Gros:2017wyj},
with a gamma-ray energy that could be tuned from 1.7 to 74 MeV.

\begin{figure}[t]
\centering
\includegraphics[scale=.125]{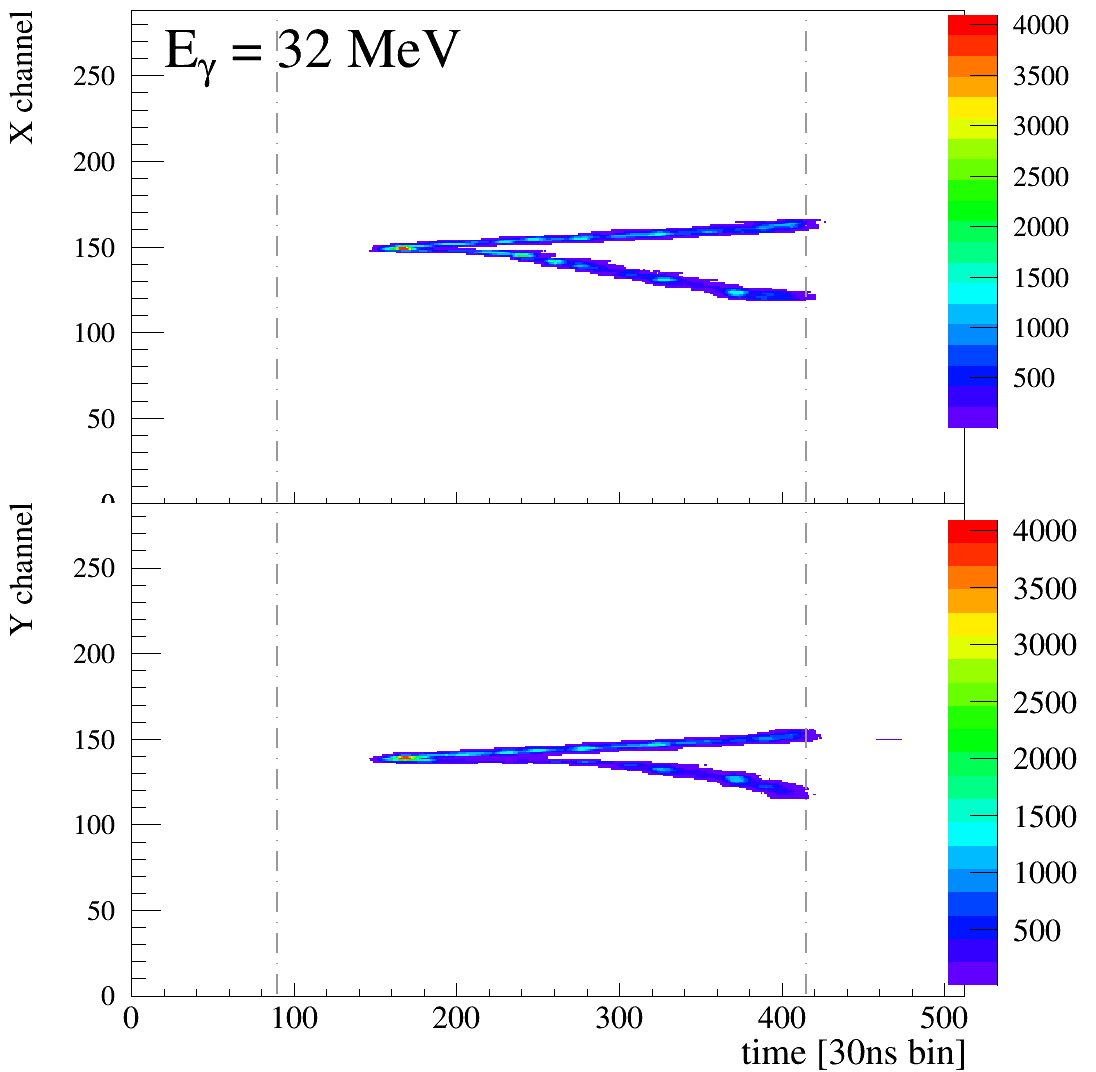}
\hfill
\includegraphics[scale=.275]{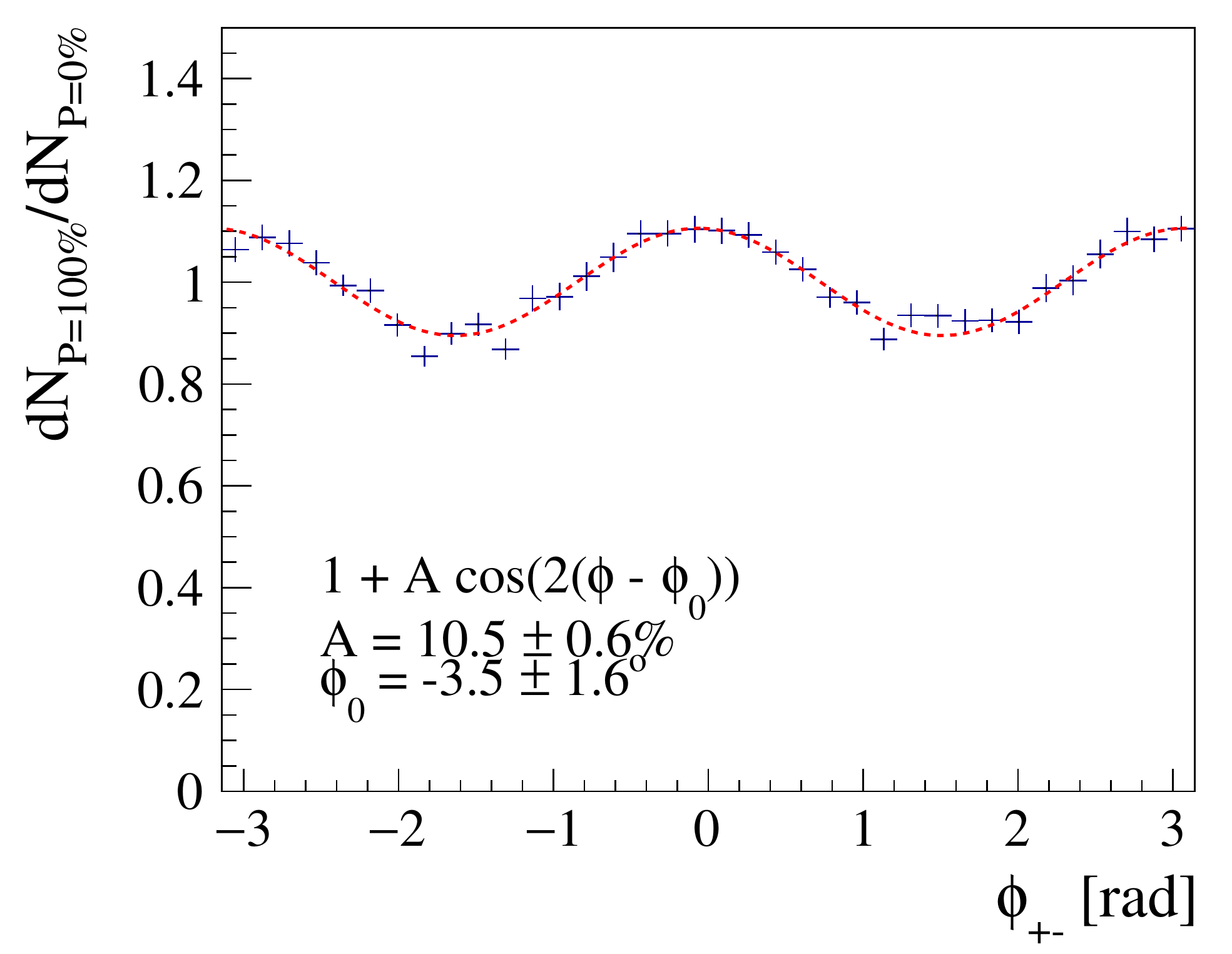}
\caption{TPC polarimeter characterization on beam:
 Left: conversion of a photon from the BL01 gamma-ray beam line at
NewSUBARU, into the $2.1\,\bbar$ argon-isobutane (95-5\,\%) gas mixture
of the HARPO TPC prototype;
the two ``maps'' are shown, that is, the two $x,t$ and $y,t$
projections of the conversion event;
the vertical dashed lines denote the physical limits of the detector,
i.e. the cathode (left) and the anode (right).
Right: distribution of the azimuthal angle of 11.8\,MeV gamma-rays
(ratio of the fully linearly polarized to the linearly non-polarized)
\citep{Gros:2017wyj} (with permission). 
\label{pair:prototype:tests}}
\end{figure}

The drift volume was cubic, $(30\,\centi\meter)^3$, filled with a
(95-5\,\%) argon-isobutane gas mixture at a pressure that was varied
from one to four bar.
Most of the data were taken at $2.1\,\bbar$, for which
a $220\,\volt\per\centi\metre$ uniform electric field
provided an electron drift velocity of 
$v_{\rm drift} \approx 3.3\,\centi\metre\per\micro\second$.
The signal was amplified by a then novel GEM$+$micromegas combination,
the performance of which was characterized carefully \citep{Gros:2014pza}.
The anode was segmented to two series of orthogonal strips ($x,y$) at
a pitch of $1\,\milli\meter$, and the signal sampled at a pace of
$30~\nano\second$, equivalent to $1\,\milli\meter$ given the value of
the drift velocity.
The prototype was routinely exposed to beam for five weeks, with an incoming
single-track background rate of several kHz, that is larger than that
expected on a space mission, a background that created negligible pile-up and no
visible gas degradation in a sealed configuration.
An excellent value of the polarization-asymmetry dilution was observed
(Fig. \ref{pair:prototype:tests} and \citep{Gros:2017wyj}).

\subsection{AdEPT}
\label{sec:adept}
 
The Advanced Energetic Pair Telescope (AdEPT) is currently being
developed at NASA/GSFC.
The primary science goal of AdEPT is to study the polarization of
photons with energies above 1 MeV, interacting via pair production
\citep{Hunter:2014}.
The polarization angle of the photon is observable as the azimuthal
angle of the electron-positron pair.
To preserve the azimuthal angle of the pair, and subsequently measure it, 
requires that Coulomb scattering of the pair in
the interaction medium be minimized. Thus, the density of the medium must be low.

A low interaction medium density is easily achieved with a gaseous TPC
detector ($\rho_{Ar} = 1.782 \times 10^{-3} (P/\bbar) \gram / \centi\meter^3$).
However, the gaseous medium has a corresponding large radiation length
($X_{0,Ar} = 19.55 \gram / \centi\meter^2
 = 1.1 \times 10^4 (1\,\bbar/P) \centi\meter$).
Maximizing the interaction probability
($\mu_{nuclear}(E = 10\,\mega\electronvolt) = 1.03 \times 10^{-2} 
 \centi\meter^2/\gram$) requires increasing the
overall instrument interaction depth to compensate for the low
interaction medium density.
However, the maximum interaction depth is limited by the maximum
allowable diffusion of the ionization electrons as they drift through
the TPC. The diffusion of the ionization electrons obscures the tracks and
consequently reconstruction of the azimuth angle of the
electron-positron pair. A useful limit to the maximum TPC drift distance is that 
distance which results in an electron 
diffusion equal to twice the pitch of the readout plane
\citep{Arogancia:2007pt}.
Increasing the total detector depth beyond this limit requires adding
additional readout planes or reducing the diffusion of the drifting
ionization charge.

The NI technique, discussed above, is an effective approach to reducing 
the diffusion in a TPC. The introduction of an electronegative component 
into any TPC gas mixture effects the operation of the TPC in three ways.
One, the diffusion of the ionization charge is reduced to the thermal limit, 
two, the drift velocity of the ions, which depends on the major component of 
the gas, is reduced by 3-4
orders of magnitude compared to the drift velocity of free electrons in the gas, and 
three, the effective gas gain of the readout plane is reduced.
This reduction in gas gain, corresponding to a reduction of the path over which the 
Townsend avalanche develops, is due to the initial acceleration of the NI process 
leading to knocking the ionization electron free from the NI and allow the avalanche
to occur. To take advantage of the reduced diffusion in a NI-TPC, the 
effects of the reduced drift velocity and increased gain must be accounted for in 
the TPC design.

\paragraph{\bf Design of a NI-TPC Pair Telescope }

The NI diffusion is reduced to the thermal limit, independent of the TPC gas mixture 
\citep{Dion:2011}, however, the NI drift velocity depends on the molecular mass of the
TPC gas. The velocity in low mass gases, e.g. argon, is faster compared with 
heavier gases, e.g. xenon. The NI drift velocity, in argon is $\approx$20 m/s, 
corresponding to a total drift time of $\approx$ 50 ms/m of drift. 
This slow drift velocity allows the spatial resolution of the TPC
readout to be very high while at the same time reducing the speed
(bandpass) and power of the readout electronics.

Reduction in the diffusion from the free electron
diffusion rate, Eq. (1), to the thermal limit is dramatic. For example, at
$1 \kilo\volt/(\centi\meter \cdot \bbar)$,
the electron diffusion in a mixture of argon and CO$_2$
is reduced from 300$\micro\meter/\sqrt{\centi\meter}$
\citep{Piuz:1983} to 80$\micro\meter/\sqrt{\centi\meter}$
at 300K. This reduction allows the drift distance, or detector interaction
depth, corresponding to a given amount of diffusion, to be increased
by a factor of $\approx$16, thereby dramatically increasing the interaction
probability of the instrument without additional layers of readout
electronics.

The slow drift velocity of the electron-positron tracks in a NI-TPC
does not absolutely rule out the use of a calorimeter.
If the electron and positron from an interacting photon cross the
readout plane and enter a calorimeter directly underneath the readout
plane, their energies from a position sensitive calorimeter could be
correlated in time with the TPC signals of the electron-positron crossing the readout plane.
Requiring this pair event geometry, however, severely reduces the
effective area of the telescope and eliminates the nearly isotropic
($2\pi$ steradian) sensitivity achievable with a TPC based gamma-ray
telescope operating in low Earth orbit.

The slow drift velocity does, however, introduce timing coincidence
problems for a gamma ray telescope if a design similar to previous
gamma-ray telescopes (e.g. EGRET, {\sl Fermi}-LAT) is following.
All previous imaging telescope designs have consisted of a
multi-layered tracker system mounted above a calorimeter and the
tracker surrounded by an anti-coincidence system (ACS).

Using a NI-TPC as the interaction medium and electron-positron pair
tracker for a pair telescope and realizing the elimination of the ACS
and calorimeter subsystems from the telescope design requires
instrument solutions to overcome the slow drift velocity of the NIs.
The advantages of eliminating the ACS and calorimeter subsystems are
two-fold.

First, eliminating these two massive and complex detector subsystems
reduces telescope complexity, mass and hardware cost.
Second, the geometric size of the telescope, and hence its
sensitivity, is determined by the dimensions of the launch vehicle
payload shroud rather than total telescope mass.

We discuss the AdEPT instrument solutions being developed to these
concerns below.

\paragraph{\bf Elimination of the ACS}

Previous telescope designs utilized an ACS to veto the instrument
triggering circuit for a few $\micro\second$ after a cosmic ray
traverses the detector volume.
The slow drift velocity of a NI-TPC would require the veto signal to
last tens of milliseconds in order for the ionization charge left by a
cosmic ray traversing the volume to completely drift through the
volume.
In a 550-600 km circular low Earth orbit, the cosmic ray proton flux
is $2.6 \times 10^4$ protons/s/m$^2$ \citep{SPENVIS}.
This proton flux and the long veto signal period would result in the
TPC readout being gated off 100\% of the time.
Thus, we conclude that the ACS is ineffective and can be eliminated
from the design of a gamma-ray telescope using a NI TPC.
Since the ionization charge produced by the cosmic-rays cannot be
discriminated, the problem becomes how to locate the interacting
gamma-rays in the background of cosmic ray tracks.

Modern computer hardware, particularly graphics processing units (GPU)
and field programmable gate arrays (FPGA) have evolved to the point
where artificial intelligence/machine learning techniques can be used
to search the NI TPC data for the electron-positron pair vertex
signature of interacting gamma rays.
The raw data is read out continuously and segmented into overlapping
time intervals of about 50 ms duration.
Each interval is then searched for the inverted ``$\Lambda$''
signature of a pair interaction.
Those intervals containing an interaction are discriminated from the
large number of background tracks from cosmic ray and low-energy
photon interactions \citep{Garnett:2020}.
These background events are eliminated from the raw data and only
those intervals containing pair interactions form the instrument
science data.

\paragraph{\bf Eliminating the Calorimeter }

As discussed in Section \ref{sec:transport}, the energy loss of a charged
particle traversing a medium is described by the Bethe formula, see
eq. (33.5) of \citep{Zyla:2020zbs}.
The mean energy loss of electrons is tabulated as a function of energy
in ESTAR \citep{NIST:electrons}.

The mean energy loss is then compared to the calculated average energy loss
\citep{Berger:2005}
to determine the energy of the original electron
\citep{Allison:1974,Allison:1976,Allison:1980,Cobb:1976}.
The positron energy loss and electron energy loss differ by a few
percent \citep{Rohrlich-Carlson:1954}.
This difference, however, is too small to be used to differentiate the pair
electron from the positron.

For charged particles traversing a gaseous TPC, the actual energy
loss, in terms of ion pairs per cm, is Poisson distributed about the
mean and fluctuates widely.
Thus, any single sample of the dE/dx is a poor indication of the
particle energy.
The high granularity (sampling rate) of a TPC provides a large number
of samples of the dE/dx loss in the gas.
The mean energy loss can be determined from these samples and the
actual energy of the electron and positron can be estimated.

The energy loss as a function of particle energy is double valued
about its minimum ionization value.
Thus, any mean energy loss corresponds to two possible particle
energies, one above the minimum ionizing energy and one below.
The lower value can be confirmed or discarded based on the total range
of the particle range.
A particle with dE/dx below (dE/dx)$_{min}$ will have more pronounced
Coulomb scattering or will stop in the gas.

\paragraph{\bf AdEPT Instrument Design }

The AdEPT instrument takes advantage of the NI-TPC technique to
provide a low density interaction medium and to optimize the
sensitivity for pair polarization.
The gas mixture, 90\%/10\% argon/methane plus $\approx$40 torr of
CS$_2$, was chosen to provide, 20 m/s, drift velocity.
A total pressure of 1.5 atm was chosen to provide significant pair
interaction depth, while minimizing the Coulomb scattering of the 
electron-positron, and allows for
a thin (3 mm aluminum) pressure vessel \citep{Hunter:2018wns}.

The AdEPT instrument is composed of two gaseous NI-TPCs stacked so
their drift electrode is common to the two TPCs and the two readout
planes are at the top and bottom of the stack \citep{Hunter:2018wns}.
Each TPCs has large, 4 m$^2$, geometric area and 1 m depth and
two-dimension readout plane based on the micro-well detector (MWD)
\citep{Deines-Jones:2002}.
This configuration of the TPCs allows the largest physical separation
between the common high voltage drift electrode and the lower voltage 
readout planes, spacecraft
structures and pressure vessel, which are at ground potential.

The two-dimension MWD readout plane has electrodes in both the X- and
Y-dimensions on $400 \micro\meter$ pitch.
The readout sampling frequency, 20 m/s / 400 $\micro\meter$ = 50 kHz,
is chosen so that the vertical coordinate digitization is also 400
$\micro\meter$.
This results in a very high granularity of $\approx$25 samples per cm of track
for the AdEPT NI TPC and only $\approx$4 cm segment of the electron or
positron track provides $\approx$100 samples of the dE/dx energy loss of the
particle.

\paragraph{\bf Micro-Well Detector Readout Plane }

The MWD readout plane is fabricated using flex printed circuit
board techniques.
The currently available flex circuit technology allows for fabrication
of multi-layer circuits with line and gap separations as small as 75
$\micro\meter$ on areas up to about 50 $\times$ 50 cm$^2$.
This allows for large area TPC readouts to be implemented by tiling
detector smaller pieces together.
The only limitation to the readout dimension and area is the
capacitance of the electrodes.

The micro-well detector is fabricated as a two layer flex circuit with
subsequent laser ablation to form the micro-wells
\citep{Deines-Jones:2002}.
The cathode electrode strips, about 200 $\micro\meter$ wide on 400 $\micro\meter$ pitch, are on the top surface of the flex circuit and the orthogonal anode electrode strips, similarly about 200 $\micro\meter$ wide on 400 $\micro\meter$ pitch, are on the lower layer.
The cathode electrodes are perforated with 150 $\micro\meter$ diameter openings aligned with the crossings of the underlying anode electrodes.
A subsequent operation, using excimer laser ablation, is used to open the 100 $\micro\meter$ diameter micro-wells in the center of each hole in the cathode electrode.
The thickness of the insulating layer determines the depth of the
wells and the maximum gas gain \citep{Deines-Jones:2002}.

This MWD geometry was considered to be a promising detector technology since the avalanche occurs in the strong gradient of the electric field in the wells and the charge is collected onto the anodes.
Motion of the positive ions resulting from the avalanche induce an equal but opposite signal on the cathodes.
Fabrication of large area MWDs using laser ablation coupled with real-time optical alignment to insure that the wells were centered on the cathode openings was investigated over many years at GSFC ultimately producing detectors as large as 30 $\times$ 30 cm$^2$.
These detectors were operated with gas gain as high as $10^4$ \citep{Son:2010}.

A wide range of laser and chemical etching techniques were explored with industrial vendors to transfer the MWD to industrial fabrication.
However, we realized that the position of the micro-wells had to be concentric, within a few microns, of the circular openings in the cathode electrodes. This accuracy requirement rendered industrial fabrication impractical.
Deviation of the well from being centered in the cathode holes, by more than a few $\micro\meter$, led to breakdown of the anode-cathode voltage which limited the gas gain performance.
The high gain lifetime of these detectors was also decreased by reduced breakdown voltage.
This effect, attributed to charge migration in the insulating layer caused by the very strong electric field in the wells, limited the lifetime of these detectors to only a few months of operation.
These considerations led us to discontinue development of MWDs at GSFC. 

\paragraph{\bf The $\micro$-PIC Detector Readout Plane }

We are currently updating the design of the AdEPT instrument,
retaining the NI TPC concept and size, but replacing the MWD readout
with the Japanese $\micro$-PIC+GEM detector developed for the ETCC
discussed in Section \ref{sec:etcc} \citep{Takada2011}.
A schematic diagram of the $\micro$-PIC, without the GEM layer, is
shown in Figure \ref{fig10}.

The electrode structure of the $\micro$-PIC detector is similar to the MWD and is fabricated as a multi-layer flex circuit.
The cathode electrodes, on the top surface, are separated from the anodes on an underlying layer with a polyimide insulating layer.
The anode electrodes have posts (filled vias) extending up through the insulating layer, typically about $\approx$100 $\micro\meter$ thick, and exposed in the center of vias in the cathodes.
These exposed anode posts and surrounding cathode strips form the proportional detector structures.
The anode-cathode gap is $\approx$100 $\micro\meter$ and the Townsend avalanche forms in the gas above the cathode layer.

\paragraph{\bf Readout Electrode Design }

The front-end electronics of the readout plane, typically a charge sensitive amplifier on each electrode, can be implemented either with discrete electronics or as an ASIC.
These implementations have an intrinsic minimum equivalent RMS noise value of about $10^3$ electrons.
Thus, a reasonable minimum detector gain for a non-NI TPC is
$\approx 3 \times 10^3$ to provide a minimum three-to-one signal-to-noise ratio.
Implementation of the NI technique reduces the effective gain of the readout because the NI must move deeper into the strong electric field until the ionization electron is knocked off of the NI at which point the free electron begins the avalanche process.
This motion of the NI reduces the effective gain of the TPC readout by a factor of $\approx$200.
The minimum gain of a NI TPC must be increased by this factor to retain the minimum signal-to-noise ratio.
Thus, the required minimum TPC gain is
$200 \times 3 \times 10^3 = 6 \times 10^5$
which is higher than is achievable with a single micromegas amplification stage \citep{Veenhof:2010}.
The solution, for a NI TPC, is an additional gain stage in the form of a gas electron multiplier (GEM).

The anode and cathode signals of a TPC are negative and positive charge, respectively.
Thus the charge amplifier must have bi-polar response or two different versions of the amplifier are needed for the anodes and cathodes.
These amplifiers can be implemented using operational amplifiers and other discrete components or an application specific integrated circuits (ASIC) can be used.
The discrete approach offers a means to quickly develop a modest number of channels when beginning detector development.
This approach is not the lowest power solution, in most cases, but does allow for easy modification of parameters such as gain and shaping time.

The readout electrodes, which operate at high voltage, require a high voltage blocking capacitor to couple the charge signal to the amplifier inputs.
These capacitors are large compared to the detector pitch and are a design problem for the electronics.
An option, explored for AdEPT instrument is to float the amplifier grounds to the corresponding anode or cathode voltages thereby eliminating the need for high voltage blocking capacitors.
This approach also requires that the grounds of all the subsequent analogue electronic stages be floated.
This is done up to the digitizer stage where the serial output of which is easily level shifted and the remainder of the event selection and background discrimination is done at spacecraft ground. 

\paragraph{\bf ASIC Readout Electronics }

The advantages of discrete electronics are quickly lost when the channel count exceeds a few hundred channels at which point selection of an integrated circuit becomes appealing.
Over the past several years there have been many general purpose ASICs developed for large experiments at CERN and by industry, e.g. IDEAS \citep{IDEAS}.

The design of an ASIC unique to a specific detector is another option that offers the advantages of tailoring the electronics to the detector and incorporating specific features into the electronics.
Such development requires several years of lead time, is expensive, and cannot typically be done until the detector performance is well defined.

An ASIC design investigated for AdEPT is a bi-directional switched capacitor charge integrating amplifier.
The adjustable integration time, 20-15 $\micro\second$, provided direct digitization of the TPC charge for each electrode with $400 \,\micro\meter$ resolution.
The advantage of integrating the charge in the front-end electronics is that the discrete charge density corresponding to a discrete TPC voxel (three-dimensional volume) is directly measured.
This ASIC design was completed and the performance simulated, however, funding issues prevented fabrication of test ASICs. 

\paragraph{\bf AdEPT Performance }

Expected performance of the AdEPT gamma-ray polarimeter based on a NI TPC in low Earth orbit is summarized below.
The detailed calculations are given in \citep{Hunter:2014}.

The effective area compared to the {\sl Fermi} LAT front is only $\approx$20\% at
100 MeV but increases rapidly at lower energies and is about five
times higher than {\sl Fermi} LAT at 20 MeV.
The point spread function approaches the kinematic limit for pair
production (the angular resolution due to the unmeasurable nuclear
recoil) and is less than twice the kinetic limit \citep{Bernard:2012uf}
up to $\approx$150 MeV, where $\theta_{68} \approx 0.6 \degree$.
This excellent angular resolution contributes to a continuum
sensitivity of about 10 mCrab over the energy range from 5 to 200 MeV.
The minimum sensitivity,
$2 \times 10^{-6} \, \mega\electronvolt \, \centi\meter^{-2} \, \second^{-1}$,
is reached at 70 MeV.
The minimum detectable polarization (MDP) for a 10 mCrab source is
$\approx$4\% and for a brighter, 100 mCrab source, the MDP decreases
to $\approx$0.7\% at 15 MeV.

\paragraph{\bf AdEPT Future Prospects }

The next phase of the AdEPT mission development will assemble a prototype of the AdEPT TPC using the $\micro$-PIC+GEM readout plane and existing discrete electronics.
Testing of this prototype with radioactive sources will provide data with realistic level of backgrounds.
This data will be used to develop the software solutions to discriminate the electron-positron pair interactions from the background and determine the electron-positron momenta.
Additional software will determine the gamma-ray incident direction, energy and polarization angle from the pair momenta.

\subsection{Liquid or Solid TPCs.} 
\label{subsec:liquid:solid}

Given the volume limitation on a space mission and the weight of the
pressure vessel needed for a gas detector, using a liquid detector
seems to be tempting \citep{Caliandro:2013kba}, with a density gain of
a factor of 840 for Argon, with respect to $1\,\bbar$ gas.
For polarimetry, the performance of the tracking system must scale
accordingly, something which presents several difficulties.
In particular, the typical collected electrical charge is left
unchanged by the double scaling (same collected charge on a 1\,cm
track segment in $1\,\bbar$ gas as on a $12\,\micro\meter$ segment in
a liquid):
an issue arises from the fact that gas detectors allow charge
amplification in the gas while liquid argon does not,
so single-phase liquid-argon detectors do not allow such a
small scale tracking.

Some amplification can be achieved, 
on Earth, by the use of a
two-phase system in which the active target consists of liquid,
from which the electrons are extracted for amplification in the gas.
In zero gravity on orbit, the stability of the gas-liquid interface
might be an issue, possibility solved by performing amplification in
bubbles inside the holes of GEMs \citep{Erdal:2019wfb}.
An other possibility to fix the dense-gas interface is to use a solid
TPC, as the electron-transport properties of solid noble gases are
similar to that of liquid noble gases
(though with a somewhat larger drift velocity)
\citep{Aprile:1985xz}.

An other issue is diffusion, as the coefficient saturates in
liquid argon for high electric field, at a value of 
$d \approx 100\, \micro\meter/\sqrt{\centi\meter}$.
The diffusion of the electron cloud during drift cannot 
scale with density, and the two tracks merge to a common blurred
single track close to the vertex.
Polarimetry with conversions to pairs, with a dense (liquid 
or solid) TPC, seems to be out of reach.

The Compton Spectrometer and Imager (COSI) project is using a set of
1.5-centimeter thick Germanium slabs as a combined
(active-target $+$ calorimeter) telescope, read with a two-fold series
of strips; as the interaction depth in the detector is inferred from
the charge collection time difference between the two sides
\citep{Kierans:2019aqz}, COSI can rightfully be described as a
Germanium TPC.

~

More on dense-phase TPCs and their possible use in gamma-ray
astronomy is presented in section \ref{sec:dense}.

\subsection{Effective area}
\label{subsec:effective:area}

In contrast to ``thick'' active targets, in which the conversion
probability of the photon is close to unity and the effective area is
the product of the geometric area by the efficiency, here for ``thin''
active targets, it is better expressed as the product of the detector
sensitive mass, $M$, by the attenuation coefficient, $H$, which is 
found to be approximately proportional to $Z^2/A$
(atomic number and mass, respectively) \citep{NIST:gamma}.
$H(E)$ tends asymptotically, at high energy, to $7 / (9 \,X_0)$.
The number of events is 
\begin{equation}
 N = t M \int \epsilon(E) f(E) H(E) \dd E 
 \label{eq:nombre:evts}
\end{equation}
$t$ mission duration; $\epsilon(E)$ efficiency.
For a ``Crab-like'' source with a spectral index of $\Gamma = 2$,
with flux $f(E) = F_0 / E^2$,
with
$F_0 = 10^{-3} \mega\electronvolt \centi\meter^{-2} \second^{-1} $
and perfect efficiency,
\begin{equation}
 N = t M F_0 \int \gfrac{H(E)}{E^2} \dd E 
 \label{eq:Crab}
\end{equation}

Values of $\displaystyle \int \gfrac{H(E)}{E^2} \dd E $ for some
materials commonly used in gamma-ray telescopes are listed in
Table~\ref{tab:prop}.
The variation of $f(E) H(E)$ as a function of $E$ is available in Fig. 
2 of \citep{Bernard:2013jea}.

\begin{table}[!h] \small
\begin{tabular}{lllllllllllllll}
 & Ne & Si & Ar & Ge & Xe \\
 $Z$ & 10 & 14 & 18 & 32 & 54 & & atomic number
 \\
 $A$ & 20.2 & 28.1 & 40.0 & 72.6 & 131. & & mass number
 \\
 $X_0 \rho$ & 28.9 & 21.8 & 19.5 & 12.2 & 8.48 & $\gram/\centi\meter^2$ & specific radiation length
 \\
 $H(100\,\mega\electronvolt)$ & 17.0 & 23.5 & 27.0 & 45.3 & 67.1 & $\centi\meter^2/\kilo\gram$ & photon attenuation coefficient @ $100\,\mega\electronvolt$ 
 \\ 
 $\displaystyle \int \gfrac{H(E)}{E^2} \dd E $ & 1.90 & 2.67 & 3.17 & 5.32 & 8.28 & $\centi\meter^2 / (\kilo\gram \, \mega \electronvolt)$
 \\
$F_0 \displaystyle \int \gfrac{H(E)}{E^2} \dd E $ & 60 & 84 & 100 & 168 & 261 & $10^3 (\kilo\gram ~ \annee)^{-1} $ & number of (thousands) events per kg per year, $\epsilon =1$
 \\
 $E_{1/2}$ & 9.8 & 9.7 & 9.3 & 9.4 & 8.8 & $\mega \electronvolt$
\end{tabular}
\caption{Properties of some material.
 See text. \label{tab:prop}}
\end{table}

For a $1\,\kilo\gram\cdot\annee$ argon-mission with full efficiency,
acceptance, exposure and perfect dilution down to threshold,
($\epsilon=1$),
and $D=1$, we would observe $N \approx 10^5$ events with an average
polarization asymmetry of $\langle A \rangle\approx 0.33$ and,
therefore, a precision of the measurement of $P$, for small $P$, of
$\sigma_P \approx\gfrac{1}{\langle A \rangle} \sqrt{\gfrac{2}{N}} \approx 0.0135$.
Note though, that the median energy $E_{1/2}$ above which half of the
collected data would lie is as low as
$E_{1/2} \approx 10\, \mega \electronvolt$, which questions the ability to
trigger / select / reconstruct / analyze low-energy conversions in the
polarimeter.

In the more realistic case of a $1\,\kilo\gram\cdot\annee$
argon-mission with a $10\,\mega\electronvolt$ threshold,
$\epsilon = 0.1$ above threshold and
$D=0.5$ dilution, 
we would obtain $N\approx 5000$ events with 
$\langle A \rangle\approx 0.232$,
$\langle A_\text{eff} \rangle\approx 0.116$,
and
$\sigma_P \approx 0.17$. 
Therefore the polarimetry of a cosmic source with pairs should focus
on the brightest sources of the MeV sky, in the first place, and with
a good-dilution, low-threshold, large-sensitive mass detector in orbit
for several years.

\subsection{Angular resolution}
\label{subsec:angular:resolution}

The primary interest of developing a telescope with an active target
consisting in a gas TPC is the excellent angular resolution and,
consequently, a sizeable sensitivity to the linear
polarization of the incoming radiation.
After a photon of momentum $\vec{k}$ converted to a pair of leptons
with momenta
$\vec{p_+}$ for the positron,
 $\vec{p_-}$ for the electron and
 $\vec{q}$ for the recoiling nucleus, with
 $\vec{k} = \vec{p_+} + \vec{p_-} + \vec{q}$, 
we want to reconstruct the direction of the candidate photon from
the measured values of $\vec{p_+}$ and $ \vec{p_-}$, as in general the
track of the recoiling nucleus cannot be reconstructed.
Therefore the single-photon angular resolution receives contributions
from the following sources:
\begin{itemize}
\item (1) the missing recoil momentum;
\item (2) the uncertainty in the magnitude of the momentum of each lepton.
\item (3) the uncertainty in the direction of the tracks;
\end{itemize}

(1) The kinematic limits for the recoil momentum are extremely large,
for nuclear conversion from 
$q_m = k - \sqrt{k^2 -4 m^2} \approx 2 m^2/ k$ to
$q_M = (k+\sqrt{k^2 - 4 m^2}) (k+M) / (2k+M)$,
where $M$ is the nucleus mass, 
but in practice the high ($q \gg mc$) part of the $q$ spectrum is
strongly suppressed, both by the presence of a $1/q^4$ factor and of
$1/(E_- - p_- \cos \theta_-)$ and
$1/(E_+ + p_+ \cos \theta_+)$ terms in the differential cross section
(the recoil direction is mainly transverse to the direction of the
photon, and all the more so, asymptotically, at high energy).
The induced contribution to angular resolution at 68\,\% containment
is found to be
\citep{Bernard:2012uf,Bernard:2019znc}
\begin{equation}
 \theta_{68}\approx b \times {E}^{-5/4},
 \quad \text{with} \quad
 b = 1.5\,\radian\, \mega\electronvolt^{5/4}
\end{equation}

As the angular kick is not Gaussian distributed, 95\,\% and
99.7\,\%-containment values might be of interest too, see
\citep{Gros:2016zst,Bernard:2019znc}.
 
(2) The contribution from the 
uncertainty in the magnitude of the momentum of each lepton
varies like $1/E$, as the fraction of the photon energy carried away
by the positron, $x_+ \equiv E_+ / E$ has a distribution that has a
mild variation with energy, and as the distributions of the polar
angles of the electron and of the positron, $\theta_+$ and $\theta_-$,
scale like $1/E$.
The relative track momentum precision, $\sigma_p / p$, depends on the
device that is used for the measurement and is not detailed here.
For $\sigma_p / p = 10\,\%$, the induced contribution on the
single-photon angular resolution is found to be negligible, compared
to the two other contributions (Fig. 6 of \citep{Bernard:2012uf}).

(3) Tracking, the determination of the direction of the tracks, faces
the limited spatial resolution of the detector and the deleterious
effect of multiple scattering.
A tracking method that takes into account the two effects in an
optimal way is the Kalman filter \citep{Fruhwirth:1987fm}.
\begin{itemize}
\item
For continuous media, and for momenta for which multiple scattering is
sizeable, and under the hypothesis that the
detector is thin enough that momentum stays nearly constant throughout
propagation, the precision of the measurement of the track angle at
``origin'' is (eq. (1) of \citep{Bernard:2019znc})

\begin{equation} 
 \sigma_{\theta} =
 \gfrac{\sigma}{l}
 \sqrt{
 \frac
 {2 \, x^3 \, \left(\sqrt{4 j - x^2} + \sqrt{- 4 j -x^2} \right)}
 { \left(\sqrt{4 j - x^2} + j x \right) \left(\sqrt{- 4 j - x^2} - j x \right)}
 }
 ,
 \label{eq:single:track}
\end{equation}

where $x$ is the longitudinal track sampling pitch, $l$, normalized to
the detector scattering length $\lambda$ \citep{Innes:1992ge},
$\sigma$ is the precision of a measurement of the track position, 
and $j$ is the imaginary unit.
\begin{itemize}
\item
At low momenta (high $x$), only the two first measurements contribute
significantly, and the angular precision is
\begin{equation}
\sigma_\theta \approx \sqrt{2} \sigma / l.
 \label{eq:single:track:highx}
\end{equation}

This takes place for 
$p < p_x \equiv p_0 \gfrac{l^{3/2}}{2^{1/3} \sigma \sqrt{X_0}}$.
Above that value, the full power of the Kalman filter is at work, and
the precision is 
\begin{equation}
\sigma_{\theta} \approx \left({p}/{p_1}\right)^{-3/4}.
 \label{eq:single:track:lowx}
\end{equation}

\item At high momenta, such that multiple scattering can be neglected,
 track-fitting turns out to be a simple linear regression, and the
 angular resolution (for no magnetic field) is \citep{Regler:2008zza}
\begin{equation}
 \sigma_\theta =
 \gfrac{2 \sigma }{l} \sqrt{\gfrac{3 }{(N-1)N(N+1)}}
\quad \approx \quad 
 \gfrac{2 \sigma }{l} \sqrt{\gfrac{3 }{N^3}}
 .
 \label{eq:single:track:Regler}
\end{equation}

This takes place for
$p> p_u \equiv p_0 \sqrt{\gfrac{l}{X_0}} \gfrac{l}{\sigma} \gfrac{N^2}{2 \times 9^{1/3}}
\quad = \quad 
p_0 \gfrac{L^2}{\sigma \sqrt{l X_0}} \gfrac{1}{2 \times 9^{1/3}} $.
\end{itemize}
\end{itemize}

For a 4-bar, argon-based TPC
($X_0 \approx 29\,\meter$, $\rho \approx 6.6 \,\kilo\gram/\meter^3$)
with $\sigma=200\,\micro\meter$ spatial resolution, 
$l= 1\,\milli\meter$ longitudinal sampling,
$N = 100$ measurements along each track on average ($L=10\,\centi\meter$),
we obtain $p_1 \approx 58\,\kilo\electronvolt/c$,
$p_x \approx 0.3\,\mega\electronvolt/c$, and
$p_u \approx 950\,\mega\electronvolt/c$:
on most of the electron momentum range relevant for TPC-based
gamma-ray telescopes, the single-track angular resolution is described
by eq. (\ref{eq:single:track:lowx}), something which translates to a
similar expression for the contribution to the single-photon angular
resolution (Sec. 3.1.1 of \citep{Bernard:2012uf}).

At $E = 100\,\mega\electronvolt$, the kinematic contribution amounts
to $ \theta_{68}\approx 0.27\degree$ and the tracking contribution to
$\sigma_\theta\approx 0.22\degree$
(the two contributions are of similar magnitude for
$p = p_l = b^2 p_1^{-3/2}$, $p_l \approx 160 \,\mega\electronvolt/c$).

\subsection{Sensitivity, gas choice}
\label{subsec:sensitivity}

Any of the noble gases is appropriate as a base for TPC and MPGD
design, from helium to xenon, with the exception of krypton, that is
generally
avoided for the radioactive decays of $^{85}$Kr would produce a
permanent background noise in the detector, at a level of
$3.7 \times 10^5$Bq per kg of natural Kr.

It could be tempting to use a high-$Z$ gas, as the photon attenuation
coefficient, and therefore the effective area per unit mass, is
proportional to $Z^2/A$, but the radiation length $X_0$ decreases, so
$p_1 \propto 1/\sqrt{X_0}$ increases, and so the single-photon
angular resolution, $\sigma_\theta \propto (p_1/p)^{3/4}$, degrades.
The sensitivity to faint sources, $s$, defined as the source flux at
significance limit,
is proportional to $\sigma_\theta / \sqrt{H(E)}$
(eq. (8) of \citep{Bernard:2012uf}).
\begin{itemize}
 \item
 Below $p_l$, with the angular resolution at the kinematic limit,
$s \propto \sqrt{A}/Z$, presenting an interest for high $Z$.

 \item
Above $p_l$, with $\sigma_\theta \propto (p_1/p)^{3/4} \propto X_0^{-3/8}$,
and $H(E) \propto 1/X_0$,
the sensitivity becomes $\propto X_0^{-1/8}$, presenting a
mild preference for low $Z$.
\end{itemize}

For polarimetry, an analysis that is so badly demanding for high
statistics, one would favor high $Z$ too, but the dilution degrades
for higher values of $p_1$ (Fig. 20 of \citep{Bernard:2013jea}).
At low pressures (several bar), statistics is the dominant factor
and high
$Z$ is preferred, while for densities tending towards that of the liquid
(hundreds of bar) it is the opposite (Fig. 22 of \citep{Bernard:2013jea}).

\section{Dense phase TPCs}
\label{sec:dense}

The use of noble gases in a dense phase, that is, liquid or solid, is
similar to that in the gas phase, but a number of differences must be
mentioned:

\begin{itemize}
\item Most often pure noble gases are used, i.e., without a
 quencher\footnote{Note though that saturation value of the electron
 drift velocity in liquid noble gases can be enlarged by the
 addition of $\approx 1\%$ in concentration of various alkanes
 \citep{Yoshino:1976zz}.},
 so diffusion could be an issue; these pure materials are
 transparent to their own scintillation light, so the scintillation
 signal can be used to generate a trigger, and / or to measure the energy
 deposited in the material.

\item Upon ionization of the material, a large electric charge is
 generated in a small volume (electrons and positive ions), after
 which recombination takes place, something that affects both the
 efficiencies of charge collection and of the detection of
 scintillation.
Recombination can be mitigated, to some extent, by a swift separation
of the positive and negative charges, in an electric field.
The fractions of the energy of the incident particle that ends up in
ionization and in scintillation show an important fluctuation, but
they are strongly anti-correlated, so a combined analysis enables a
precision measurement of the total deposited energy.
An energy resolution of 1.7\,\% could be obtained in that way with a
LXe (liquid xenon) detector with 662\,keV photons
at $1\,\kilo\volt/\centi\meter$,
for example \citep{Aprile:2007qd}.
Scintillation and ionization yields reported in Table \ref{tab:liq}
were obtained with incident electrons.

\item Note that charge amplification in the liquid is either difficult
 (Xe, with maximum gains of a couple of hundreds) or impossible (Ar),
 as it needs to operate close to breakdown voltage.
\end{itemize}

\begin{table}
 \caption{Properties of the dense (liquid or solid) phases of noble gases, used in TPCs.
 Values
 of the radiation lengths are from \citep{Zyla:2020zbs},
 of the scintillation yields from \citep{Michniak} and \citep{Doke:2002oab},
 of the refractive indices from \citep{Grace:2015yta},
 of the ionization yields from \citep{Chepel:2012sj} and \citep{Guarise:2020ojj},
 of the electron drift velocities from \citep{Aprile:1985xz}\citep{Yoo:2015yza} \citep{Sakai},
 of the low-field mobilities from \citep{Sakai} and \citep{Miller:1968zza}.
 \label{tab:liq}}
\small
\begin{tabular}{ |l|l|l|l|l|l|l|l|l|l|l|l| }
 \hline
 & &
 \multicolumn{2}{|c|}{Ne}&
 \multicolumn{2}{|c|}{Ar}&
 \multicolumn{2}{|c|}{Xe} \\
 \hline
 & & Liq. & Sol. & Liq. & Sol. & Liq. & Sol. \\
 $T$ (boil @ 1 atm, melt) & K & 27.07 & 24.56 & 87.30 & 83.79 & 165.1 &161.4 \\
 $\rho$ & $\gram \, \centi\meter^{-3}$ & 1.204 & 1.444 & 1.396 & 1.623 & 2.953 & 3.41\\
 $X_0$ & $\centi\meter$ & 24.03 & 20.0 & 14.00 & 12.0 & 2.872 & 2.49\\
 \hline
 \multicolumn{8}{|l|}{{\bf Scintillation}} \\
 \hline
 yield & $\gamma$ / keV & 74 & 170 & 40 & & 42 & id Liq. \\
 $\lambda_{\scint}$ & nm & 80 & & 128 & & 178 & \\
 $n$ @ $\lambda_{\scint}$ & & & & $1.45 \pm 0.07$ & $1.50 \pm 0.07$ & $1.69 \pm 0.04$ & $1.81 \pm 0.03$ \\ 
 \hline
 \multicolumn{8}{|l|}{{\bf Electron transport}} \\
 \hline
 ionization yield & $e^- / \kilo\electronvolt$ & & 46 & 40 & 42 & 44 & 64 
\\
$e^-$ drift velocity $v$ at 1\,kV/cm & $\centi\meter / \micro\second$ & & & 0.225 & 0.38 & 0.2 & 0.4
\\
$e^-$ drift velocity $v$ at saturation & $\centi\meter / \micro\second$ & & 1.9 & 1.0 & 1.07 & 0.30 & 0.56 
\\
low field $e^-$ drift mobility $\mu$ & & & 600 & 475 & 1000 & 2200 & 4500 
\\
$D_T$ & $\centi\meter^2 / \second$ at $1\,\kilo\volt/\centi\meter$ & & & 15 & & 60 -- 80 & \\ 
\hline
\end{tabular}
\end{table}

\begin{itemize}
\item Note that electrons cannot drift freely in liquid neon, as they
 attract a group of atoms that make an heavy negative-charge,
 low-mobility ``bubble''.
 But solid neon is of interest, still. 
\item
 Scintillation takes place in the vacuum ultraviolet
 (see Fig. 3.27 of \citep{Aprile:Book}), so light
collection is performed through VUV transparent windows or after
 wavelength-shifting (WLS) inside the experimental vessel.
 The spectra are similar for the gas, for the liquid and for the
 solid phases (see Fig. 3.32 of \citep{Aprile:Book}),
 with typical width (FWHM) of $\approx 10 \, \nano\meter$.

\item
 Refraction indices are computed by \citep{Grace:2015yta} at the
 scintillation wavelength, from data collected at higher wavelengths.

\item
 The ionization electron yield depends on the
 experimental conditions, in particular on
the nature and the energy of the incident particle, and on
 the value of the applied
 electric field that mitigates the loss due to electron-ion
 recombination
 (solids: Table 1 of \citep{Guarise:2020ojj};
 liquids: Table 1 of \citep{Chepel:2012sj}).

\item
 The electron drift velocity depends on the temperature, especially for solids
 (See Fig. 3.11 of \citep{Aprile:Book} for Argon).
 The values of the electron drift mobility are provided close to the
 triple point \citep{Miller:1968zza}.

\item Diffusion of the drifting electrons is larger than the thermal
 limit over most of the electric field practical range, with larger
 values for Xe than for Ar (See Fig. 3.12 of \citep{Aprile:Book}) and
 for transverse diffusion than for longitudinal
 ($D_L / D_T \approx 0.1$ for Xe), see the discussion in section 4.3.1 of 
\citep{Chepel:2012sj}.
For typical values of $D_T = 25 \, \centi\meter^2 / \second$ and
 $v = 0.5 \, \centi\meter / \micro\second$
 the RMS spread is $\sqrt{2 D_T / v} \approx 100\,\micro\meter / \sqrt{ \centi\meter}$.
\end{itemize}

High-mobility liquid hydrocarbons have been considered in the past, so
as to avoid the complexity of a cryogenic apparatus.
Also the drift velocity saturates at a much larger electric field and
at a much larger value than for liquid noble gases \citep{sowada}
so the diffusion spread can be extremely small
\citep{BakaleBeck}.
Alkanes can be used and higher effective-$Z$ materials can be obtained
by replacing the central carbon atom by a heavier atom, such as for 
tetramethylsilane (CH$_3$)$_4$Si, tetramethylgermane (CH$_3$)$_4$Ge
...
Unfortunately, these materials are opaque to their own scintillation.

\subsection{LXeGRIT}

\begin{figure}
 \includegraphics[width=0.48\linewidth]{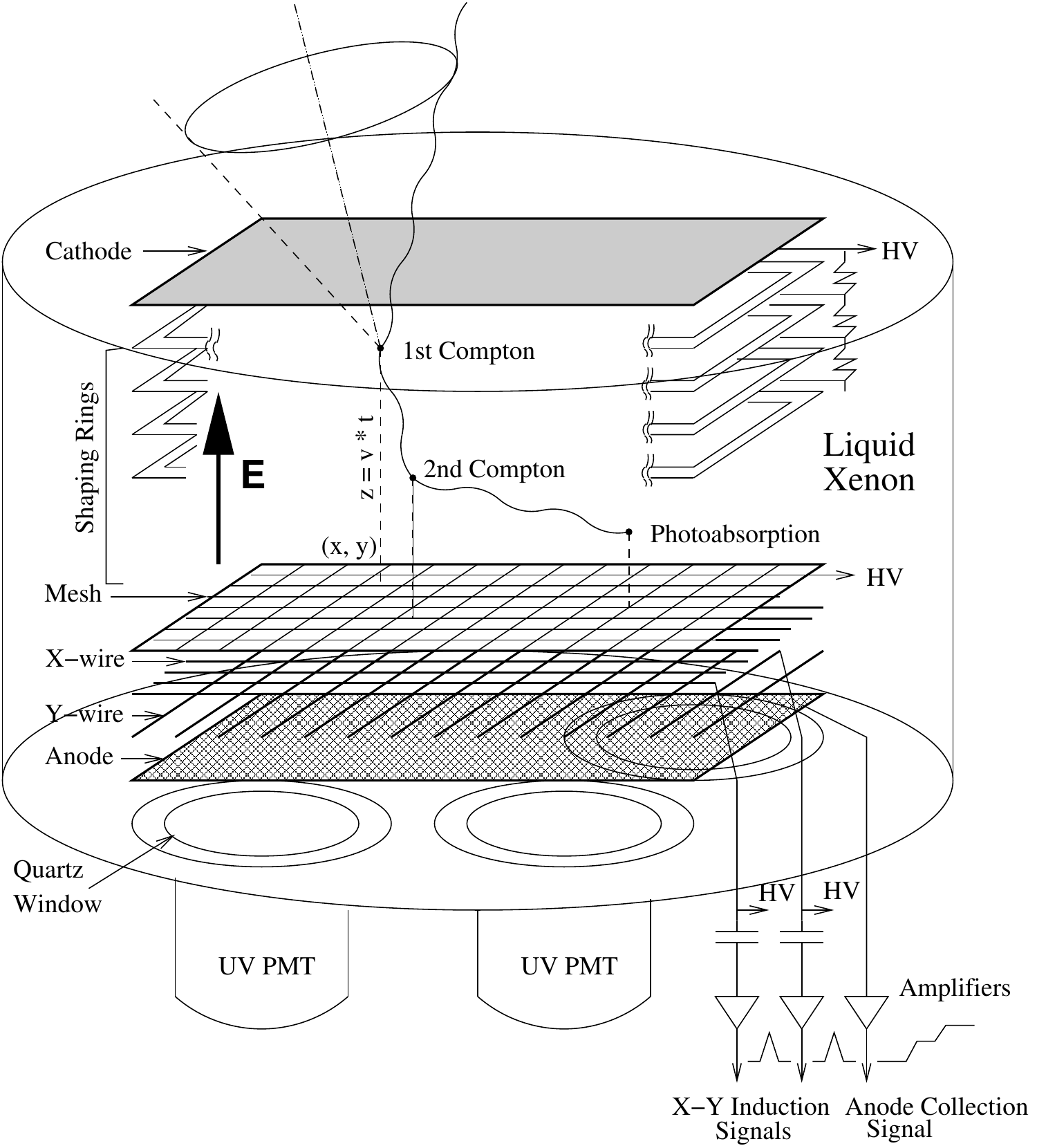}
 \hfill
 \includegraphics[width=0.48\linewidth]{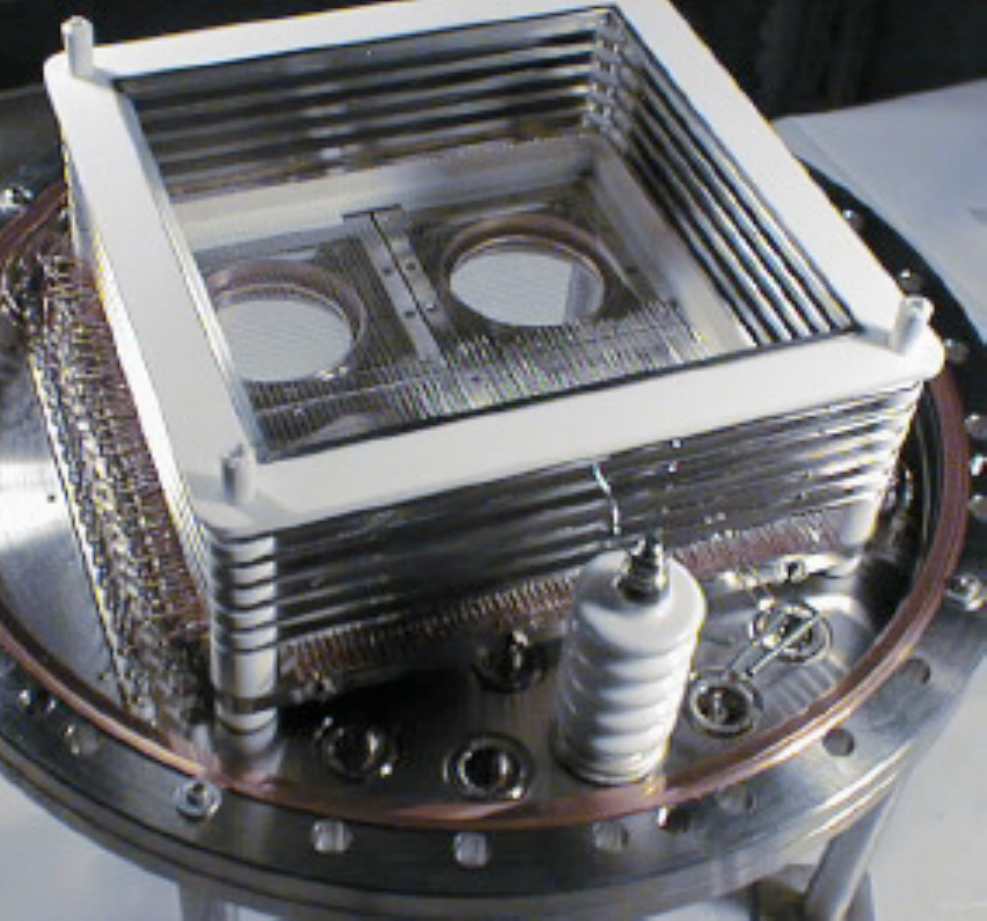}
 \caption{
Schema (left, \citep{APRILE1998425})
and picture of the TPC structure (right \citep{Aprile:2009dv})
 of the LXeGRIT telescope.
A 2-scatter, 3-site event is represented on the schema.
 \label{fig:lxegrit}
 }
\end{figure}

LXeGRIT (Liquid Xenon Gamma-Ray Imaging Telescope) is a balloon-borne
Compton telescope based on a liquid xenon TPC for imaging cosmic
gamma-rays in the energy range $0.15 - 10\,\mega\electronvolt$
with a FoV at FWHM of $\approx 1\,\steradian$.
The active target has a $20 \times 20 \, \centi\meter^2$ sensitive area,
a $7\, \centi\meter$ thickness and contains high-purity liquid
xenon at a temperature of $-95\,\degree\Celsius$, immersed in a
$1\,\kilo\volt/\centi\meter$ electric field.
The full drift duration is of $35\,\micro\second$.
The scintillation light is collected by a set of four PMTs located
below the sensitive volume through quartz windows.
After drift, the ionization electrons traverse a set of two orthogonal
wire sheets at a pitch of $3\,\milli\meter$, on which they induce a
signal, after which they are collected on a four-fold segmented anode
(Fig. \ref{fig:lxegrit}).

LXeGRIT has undergone calibration with radioactive sources, full
simulation, and a series of balloon-borne flights.
The telescope shows a
$\Delta E = 8.8 \% / \sqrt{E / \mega\electronvolt}$ FWHM energy
resolution and
 a single-interaction position resolution better
than $1\,\milli\meter$ in the transverse and than $0.3\,\milli\meter$
in the longitudinal directions, from which an ARM angular resolution of
$3.8\degree$ was obtained at $1.836 \, \mega\electronvolt$
(the energy dependence for 2- and 3- site events can be found in Fig. 14 of 
\citep{Aprile:2008ft}).

A 27 hour balloon flight was performed in 2000, of which 5 hours of
consecutive data were taken at an altitude of $39\,\kilo\meter$, or an
atmospheric thickness of $3.2\,\gram \centi\meter^{-2}$
\citep{Curioni:2002ih}.
A first level, fast trigger was formed from an OR of the PMT signals.
A second level trigger, including a cut on the number of wire hits,
enabled a rejection of single-site and charged cosmic-ray background
noise events.
The background rate in flight was actually found to be larger than
anticipated, so it was decided to reduce the trigger efficiency so as
to mitigate the data-acquisition bottleneck \citep{Curioni:2007rb}.
The energy spectrum (Fig. 9 of \citep{Curioni:2002ih}) shows a
continuum, plus a 1.46\,MeV peak due to $^{40}$K radioactivity, from
the potassium present in the ceramic part of the
structure of the detector.
The Crab nebula/pulsar has been within the field of view for several
hours of that flight; it has been estimated that hints of a detection
were within reach at the $2 - 3 \, \sigma$ significance level
\citep{Aprile;NAR}.

\subsection{Liquid TPCs as high-resolution homogeneous calorimeters}

Homogeneous liquid xenon calorimeters have been considered as high
energy-resolution active targets for gamma-ray lines in the GeV energy
range \citep{Doke:1988mt,Okada:2000tz,Baranov:1990mi}.
The angular resolution has been estimated to be
$\approx 1\,\milli\radian$ at 10\,GeV \citep{Seguinot:1990mu}, to be
compared to the PSF at 68\,\% (front) of the {\sl Fermi}-LAT
tracker, of $\approx 7\,\milli\radian$
(P8R3 release).

\section{Summary / Conclusions}
\label{sec:conclusions}

Gamma-ray astronomy is suffering from the lack
of high-sensitivity instruments at the frontier of the Compton and of
the pair-conversion energy ranges, and from the difficulty of performing
a polarimetry of the incoming radiation from gamma-ray sources.

On the Compton side, one of the issues is the complexity of the
analysis induced by the determination of the incoming photon direction
on a cone for one-scatter events, while on the pair side, the
degradation of the angular resolution at low energy due to multiple
scattering of the leptons in their way through the detector is the
main limitation.
In both cases, the ability of selecting signal photons and reject 
background noise is affected, the sensitivity is degraded, and in both cases
a key issue is the precision of the tracking of low-energy electrons.

On both sides of the frontier, major improvements have been achieved
since the beginning of this century, thanks to the development of
low-density high-precision active targets such as gas TPCs.
For Compton events, it was recognized and demonstrated experimentally
that the tracking of the scattered electron can be so precise that for
each event, the cone arc can be brought down to an almost isotropic
PSF.
TPC prototypes have undergone balloon test flights, during which the
detector and in particular the trigger system survived the intense
single-track background in the upper atmosphere, and cosmic photons
were observed.

For pair-conversion events, TPCs enable an excellent angular
resolution down to the kinematic limit (that is due to the
non-observation of the recoiling nucleus).
A measurement of the
linear polarization fraction and angle of the incoming radiation has
been demonstrated both by the analysis of simulated data and by the
characterization of a TPC prototype on a $\gamma$-ray beam.

Several MPGD techniques have been developed, that provide a high-gain,
high-rate, spark-resistent, low-jitter, low-ion backflow amplification
of the time-dependent signal that is flowing on the readout electrodes.
These amplification devices work in the so-called proportional mode,
far from the sparking mode of the EGRET spark chambers that is
suspected to have inflicted the radiation-induced chemistry processes that
have degraded the gas to the point that EGRET had to change it to
fresh gas by the year.
The design of low-ageing gas detectors can benefit from the experience
gained by the experiments at the Large Hadron Collider (LHC), that have been
routinely exposed for years to much higher radiation dose rates
than for a detector in orbit.

Higher density TPCs using a liquid or solid material have
been considered but the tracking precision is an issue as charge
amplification is limited in noble-gas liquid and mitigating diffusion
during drift by the addition of a quencher is difficult, so having the
tracking pitch and precision scale down with density is not an option.
For Compton events, furthermore, the charge collection statistics is
much lower than for semi-conductor materials like Silicon or
Germanium, so the energy resolution is not as good and nuclear
spectroscopy seems to be out of reach. 

The ``vertical'' coordinate is determined from
a time-of-flight difference in the COSI detector, so the first
gamma-ray TPC in orbit might well consist of Germanium.

\section{Acknowledgement}

This study was supported by the Japan Society for the Promotion of Science
(JSPS)
Grant-in-Aid for Challenging Research Pioneering 20K20428.

\section{Cross-References} 

``Gamma-ray Polarimetry'', 
D. Bernard, T. Chattopadhyay, F. Kislat (corresponding author) and N. Produit, 
arXiv:2205.02072 [astro-ph.IM]





\section{Table of variables used in the text}

\small

\begin{tabular}{lllllllllllllll}

$A$ & mass number \\
$A$ & polarization asymmetry \\
$A_{\text{eff}}$ & polarization asymmetry, with detector effects \\
$A_{\text{eff}}$ & effective area\\
 $\alpha$ & Compton 
 scattering angle $\alpha$ between the recoil electron\\
 & and the scattered gamma ray
\\

$\beta$ & particle velocity normalized to that of light \\
$b$ & single-photon resolution-angle kinematic-limit constant & $ b = 1.5\,\radian\, \mega\electronvolt^{5/4}$ \\
$B$ & magnetic field \\
$B$ & the background flux \\

$D$ & (polarization asymmetry) dilution factor \\
$D$ & diffusion coefficient \\
$d$ & pseudo diffusion coefficient \\

$e$ & elementary electric charge \\
$E$ & particle energy \\
$E_0$ & Olsen constant & $E_0 = 1.6\,\mega\electronvolt$ \citep{Olsen1963} \\
${\cal E}$ & electric field \\
$\epsilon$ & efficiency \\
$f$ & source flux energy distribution \\
$\gamma$ & particle Lorentz factor \\

$H$ & photon attenuation coefficient & \citep{NIST:gamma} \\
$j$ & the imaginary unit \\

$k$ & Boltzmann's constant \\
$\vec{k}$ & photon momentum \\
$L$ & length \\
$l$ & longitudinal sampling along the track \\

$\lambda$ & detector scattering length & \citep{Innes:1992ge} \\
$m$ & electron mass \\
$M$ & nucleus mass \\
$M$ & detector sensitive mass \\
$\mu$ & mobility \\
$N$ & number of events \\

$p$ & pressure \\
$p$ & particle momentum \\
$p_r$ & quencher partial pressure \\
$p_0$ & multiple scattering constant & $p_0 = 13.6\,\mega\electronvolt/c$ \citep{Zyla:2020zbs} \\
$p_1$ & detector characteristic scattering momentum \\ 
$P$ & (linear) polarization fraction of light \\
$\varphi$ & azimuthal angle \\

$q$ & recoil momentum \\
 
$\rho$ & density \\

$S$ & signal intensity \\
$\sigma$ & resolution of track position single-measurement \\
$\sigma$ & electron cloud spread due to diffusion \\
 
$t$ & time \\
$T$ & temperature \\
$\theta$ & polar angle (wrt the incident photon direction) \\
$\theta_{+-}$ & pair opening angle \\
$\theta_0$ & multiple scattering RMS angle with logarithmic correction factor neglected \\
 \end{tabular}

 \begin{tabular}{lllllllllllllll}
$v$ & velocity \\
$X_0$ & detector material radiation length \\
$x_+$ & fraction of the incident photon energy carried away by the positron \\ 
$x$ & longitudinal track sampling pitch, $l$, normalized to detector scattering length $\lambda$ & $x = l/\lambda$ \\
$x$ & transverse coordinate \\
$y$ & transverse coordinate \\
$z$ & ``vertical'' position, i.e. along the electric field \\
$z$ & particle electric charge in elementary electric charge units \\
$Z$ & atomic number 
\end{tabular}

\tableofcontents
\end{document}